\documentstyle[11pt]{article}
\def\setfonts{%
\font\ssfbig=cmss10 scaled\magstephalf
\font\ssfscr=cmss8 
\font\ssfscrscr=cmss8
\newfam\ssffam
\textfont\ssffam=\ssfbig
\scriptfont\ssffam=\ssfscr
\scriptscriptfont\ssffam=\ssfscrscr
\def\ssf{\fam\ssffam}
\font\openbig=msbm10 scaled\magstephalf
\font\openscr=msbm8 
\font\openscrscr=msbm8
\newfam\openfam
\textfont\openfam=\openbig
\scriptfont\openfam=\openscr
\scriptscriptfont\openfam=\openscrscr
\def\open{\fam\openfam}
}

\makeatletter
\@addtoreset{equation}{section}
\newdimen\normalarrayskip
\newdimen\minarrayskip
\normalarrayskip\baselineskip
\minarrayskip\jot
\newif\ifold \oldtrue \def\new{\oldfalse}
\def\arraymode{\ifold\relax\else\displaystyle\fi}

\def\@arrayskip{\ifold\baselineskip\z@\lineskip\z@
  \else
  \baselineskip\minarrayskip\lineskip2\minarrayskip\fi}
\def\@arrayclassz{\ifcase \@lastchclass \@acolampacol \or
\@ampacol \or \or \or \@addamp \or
 \@acolampacol \or \@firstampfalse \@acol \fi
\edef\@preamble{\@preamble
 \ifcase \@chnum
  \hfil$\relax\arraymode\@sharp$\hfil
  \or $\relax\arraymode\@sharp$\hfil
  \or \hfil$\relax\arraymode\@sharp$\fi}}
\def\@array[#1]#2{\setbox\@arstrutbox=\hbox{\vrule
  height\arraystretch \ht\strutbox
  depth\arraystretch \dp\strutbox
  width\z@}\@mkpream{#2}\edef\@preamble{\halign \noexpand\@halignto
\bgroup \tabskip\z@ \@arstrut \@preamble \tabskip\z@ \cr}%
\let\@startpbox\@@startpbox \let\@endpbox\@@endpbox
 \if #1t\vtop \else \if#1b\vbox \else \vcenter \fi\fi
 \bgroup \let\par\relax
 \let\@sharp##\let\protect\relax
 \@arrayskip\@preamble}
\@addtoreset{equation}{section}
\makeatother

\setfonts

\def\lvm{\leavevmode\hbox to\parindent{\hfill}}
\def\req#1{(\ref{#1})}
\def\reb#1{{\bf\ref{#1}}}

\def\BE{\begin{equation}}
\def\EE{\end{equation} }
\def\BA{\begin{array}}
\def\EA{\end{array}}

\def\bar{\overline}
\def\frac#1#2{\mathchoice{{\textstyle{{#1}\over{#2}}}}{{#1\over#2}}{{#1\over#2}}{{#1\over#2}}}
\def\ket#1{\mathchoice{{\left|{#1}\right\rangle}}{|{#1}\rangle}{|{#1}\rangle}{|{#1}\rangle}}
\def\ketch#1#2{\bigl|{#2}\bigr\rangle_{\rm ch}^{(#1)}}

\def\MW{\ket}

\def\ctop{{\ssf c}}
\def\ktop{{\ssf k}}
\def\ptop{{\ssf p}}

\def\jplus{{\ssf j}^+}
\def\jminus{{\ssf j}^-}

\def\htop{{\ssf h}}
\def\ptop{{\ssf p}}
\def\theell{{\sf l}}

\def\d{\partial}

\def\N#1{N\!=\!#1}
\def\SL#1{s\ell(#1)}
\def\SSL#1#2{s\ell(#1|#2)}

\def\half{{\textstyle{1\over2}}}

\def\fourth{{\textstyle{1\over4}}}
\def\crossbox{\mathop{\rlap{\raisebox{-1pt}{$\Box$}}\times}}

\def\cA{{\cal A}}
\def\cB{{\cal B}}

\def\cE{{\cal E}}

\def\cG{{\cal G}}
\def\cH{{\cal H}}

\def\cL{{\cal L}}
\def\cN{{\cal N}}

\def\cQ{{\cal Q}}

\def\cT{{\cal T}}
\def\cU{{\cal U}}

\def\oN{{\open N}}
\def\oC{{\open C}}

\def\oZ{{\open Z}}

\def\tensor{\otimes}
\def\tilde{\widetilde}

\def\NPB{Nucl.\ Phys.\ B}
\def\PRD{Phys.\ Rev.\ D}
\def\PLB{Phys.\ Lett.\ B}
\def\MPLA{Mod.\ Phys.\ Lett.\ A}
\def\CMP{Commun.\ Math.\ Phys.}
\def\IJMPA{Int.\ J.\ Mod.\ Phys.\ A}

\def\barpsi{\bar\psi}

\def\Hplus{H^+}
\def\Hminus{H^-}

\def\tbeta{\widetilde\beta}
\def\tgamma{\widetilde\gamma}
\def\barpsi{\bar\psi}
\def\Dbarphi{\d\bar\phi}
\def\Dphi{\d\phi}
\def\DF{\d F}
\def\DU{\d U}

\def\emt{energy-momentum tensor}
\def\hw{highest-weight}

\newtheorem{lemma}{Lemma}[section]

\newtheorem{thm}[lemma]{Theorem}

\newenvironment{rem}{%
\smallskip\stepcounter{lemma}\noindent{\bf Remark~\thelemma} \ }%
{\par\smallskip}

\newcounter{work}

\begin{document}


\thispagestyle{empty}
\hfuzz=1pt

\begin{flushright}
{\tt hep-th/9610084}\\
HUB-EP-96/52
\end{flushright}
\thispagestyle{empty}

\begin{center}
{\Large{\sc Verma Modules, Extremal Vectors, and Singular Vectors}}\\[4pt]
{\Large{\sc on the Non-Critical $\N2$ String Worldsheet}}
\\[10pt]
{\large A.~M.~Semikhatov
}\\[8pt]
{\small\sl Institut f\"ur Physik,
Humboldt-Universit\"at zu Berlin}\\ 
{\small\it Invalidenstra\ss e 110, D-10115 Berlin, Germany,}
\\[6pt]
{\small and}\\[6pt]
{\small\sl I.E.~Tamm Theory Division,
P.N.~Lebedev Physics Institute, Russian Academy of Sciences}\\
{\small\it 53 Leninski prosp., Moscow 117924, Russia}
\end{center}
\vskip-4pt

{ \addtolength{\baselineskip}{-5pt} {\footnotesize We formulate the
general construction for singular vectors in Verma modules of the
affine $\SSL21$ superalgebra.  We then construct $\SSL21$
representations out of the fields of the non-critical $\N2$ string.
This allows us to extend naturally to $\SSL21$ several crucial
properties of the $\N2$ superconformal algebra, first of all the
construction of extremal states (an analogue of different pictures for
non-free fermions) and the spectral flow transform (which then affects
the Liouville sector).  We further evaluate the affine $\SSL21$ singular
vectors in the realization of $\SSL21$ provided by the $\N2$ string.
We establish that, with a notable exception, the respective singular
vectors are in a $2:1$ correspondence, namely two different $\SSL21$
singular vectors evaluate as an $\N2$ superconformal singular vector
(however, those singular vectors that are labelled by a pair of
positive integers get these integers transposed under the reduction).
We also analyse the `exceptional' cases, which amount to a series of
$\SSL21$ singular vectors, labelled by $r\geq1$, which do not have an
$\N2$ counterpart, and discuss the mechanism by which the multiplicity
of singular vectors becomes equal to two at certain points in the
weight spaces of both algebras.}  }

{\small
\tableofcontents
}

\addtolength{\baselineskip}{2pt}
\newpage
\setcounter{page}{1}


\section{Introduction}\lvm
String theories, taken in their worldsheet formulation, are known to
possess a number of hidden worldsheet
symmetries~\cite{[GS2],[BLNW],[BLLS],[LLS],[RSS]} and a series of
embeddings~\cite{[BV],[FoF],[OP],[BOP],[BOh],[BV-top],[OV-top]}.  One
is particularly interested~\cite{[RSS]} in relations of string
theories to the affine Lie algebras (such relations can be observed at
different levels, see e.g., ~\cite{[AGSY],[HY],[FY]}).  At the same
time, a series of embeddings found between string theories show that
lower-supersymmetric strings can be `prepared' as some special states
of the higher-supersymmetric ones.

\smallskip

The $\N2$
strings~\cite{[Ade],[Marcus],[FT],[MM],[OV23],[BV-top],[Lechtenfeld]},
whose role in the M-theory has recently been proposed in
\cite{[KM],[Mart]} (see also~\cite{[Ketov]}), are interesting also
because the non-critical $\N2$ string provides a direct realization of
the affine $\SSL21$ superalgebra on the worldsheet in the conformal
gauge~\cite{[S-sl21]}.  One can therefore expect that a number of
properties of the affine $\SSL21$ algebra would be seen in $\N2$
strings.  The most far-reaching consequences would be those concerning
the structure of physical states (the BRST cohomology), aimed at an
$\N2$ extension of the results of \cite{[LZ]} (see also~\cite{[BMP]}),
and the fusion rules. As a step towards that aim one has to consider
first the structure of the $\SSL21$ representations realized in the
non-critical $\N2$ string, and this is one of the problems that we
address in the present paper.

\smallskip

As is the case with the much better studied $\N0$ models, one is
particularly interested in those representations whose highest-weight
states allow for the existence of singular
vectors~\cite{[BPZ],[DF],[MFF],[KK]}; then, after factoring out the
submodule generated by the singular vectors, one is left, in the
Virasoro and affine $\SL2$ cases, with an irreducible
representation. As to the {\it rank-3\/} affine $\SSL21$ algebra, the
situation can be more complicated due to the presence of {\it
sub\/}singular vectors (i.e., the module obtained by factoring over
`level-1' singular vectors may not be irreducible). There has been a
constant interest in the structure and explicit constructions of
singular vectors of the various infinite-dimensional
algebras~\cite{[MFF],[BSAVir],[BdFIZ],[BS],[GP],[BWW3],[W1],[B],[BFK],[Doerr2]}.
In particular, relations between different algebras, such as,
primarily, the Hamiltonian reduction, have in some cases been shown to
extend to the respective singular vectors~\cite{[GP]}. The appearance
of the $\SSL21$ algebra in the non-critical $\N2$ string is
significant since it is by the Hamiltonian reduction of the affine
$\SSL21$ that one can obtain the $\N2$ superconformal
algebra~\cite{[BO],[BLNW],[IK]}.  Thus the construction of the affine
$\SSL21$ out of $\N2$ superconformal matter `dressed' with some free
fields (the ghosts and the Liouville) can be considered as an
`inversion' of the Hamiltonian reduction~\cite{[S-inv]}, and one may
expect that these two algebras would have `{\it related\/}'
representations and `{\it related\/}' singular vectors\,\footnote{That
the Hamiltonian reduction tends to apply nicely to singular vectors
does not follow from the first-principles; indeed, the Hamiltonian
reduction does in general destroy even the Lie-algebra structure
(recall the reductions of `higher'-rank Ka\v c--Moody algebras,
resulting in W-, not Lie, algebras), and there seem to be no general
grounds to expect that it would induce any reasonable relation between
representations. In our opinion, the possibility to `extend' the
Hamiltonian reduction to singular vectors has rather to do with the
existence of the `inversion' of the reduction~\cite{[S-inv]}.  This
raises, however, an intriguing problem of `inverting' the $\SL3\to
W_3$ Hamiltonian reduction.}.  From the correspondence between
singular vectors, a relation between the `Lian--Zuckerman' states of
the respective theories should ultimately follow, since the
Lian--Zuckerman states are related to singular vectors in the Verma
modules (even though the explicit form of that relation has not been
worked out in sufficient generality, see~\cite{[KS],[IMM]}). A similar
correspondence should also exist then between the fusion rules.

\smallskip

While neither the `Lian--Zuckerman' states for $\N2$ strings nor the
fusion rules for $\SSL21$ are yet known, singular vectors are a tool
of major importance in the analysis of representations of these
algebras.\footnote{Another, seemingly unrelated, reason to be
interested in singular vectors comes from integrable `massive'
models~\cite{[BBS],[MW]}.} In this paper, firstly, we give the general
construction for singular vectors of the affine $\SSL21$ algebra. In
doing so, we will employ the properties of the $\SSL21$ algebra that
turn out to have analogues in the case of the $\N2$ superconformal
algebra.  An important feature of the affine $\SSL21$ algebra, which
shows up in its representation theory, is the existence of a spectral
flow transform, similar (and in fact, closely related) to the one
which is known to be important in the $\N2$ superconformal
algebra~\cite{[SS],[LVW],[ST3],[Gd]} and $\N2$ (critical)
strings~\cite{[BV-top],[Lechtenfeld],[KL]}. We will show that the
spectral flow transform extends naturally from the $\N2$ `matter' to
the $\SSL21$ algebra and therefore to the noncritical $\N2$ string,
where it also affects the Liouville superpartner. Secondly, we will
reduce the problem of the analysis of $\SSL21$ singular vectors to the
$\N2$ superconformal algebra: by evaluating $\SSL21$ singular vectors
in the realization provided by the $\N2$ string, we will show that the
$\SSL21$ singular vectors are in a $2:1$ correspondence with the $\N2$
singular vectors, except for a series of $\SSL21$ singular vectors,
labelled by positive integers, which do not correspond to any $\N2$
singular vector (which suggests, in particular, that the $\SSL21$
fusion rules would have an extra series as compared with the $\N2$
fusion rules). We will also find that those $\SSL21$ singular vectors
which, in some standard nomenclature, are labelled by a pair $(r,s)$
of positive integers, map to the $(s,r)$ singular vectors of the $\N2$
algebra.  At the same time, none of the singular vectors are `lost' in
the realization of the affine $\SSL21$ provided by the non-critical
$\N2$ string: in terms of the fields of the $\N2$ string, none of the
singular vectors vanishes, and different singular vectors evaluate
differently.  All this will be done for an arbitrary (complex) level
$k\neq-1$, and, accordingly, an arbitrary $\N2$ central charge
$\ctop\neq3$. The general construction of $\SSL21$ singular vectors is
then to be applied to the case of rational $k$ similarly to how this
is done for the ordinary MFF construction~\cite{[MFF]}: for the $\SL2$
case, for example, each of the two MFF formulae produces a singular
vector as many times as there are different solutions to the equations
on the parameters (in that case, the spin and the level) that
guarantee the existence of a singular vector.

\medskip

Another algebra which emerges on the $\N2$ non-critical string
worldsheet is the $\N4$ superconformal algebra~\cite{[BLLS]}. In that
respect, the $\N2$ string repeats, with some interesting
modifications, the relations existing around the non-critical bosonic
string:
\BE\BA{rcccl}
\SL2\kern-10pt&{}&{}&{}&\kern-10pt\N2\\
{}&\rlap{\raisebox{-2pt}{\mbox{\LARGE $\searrow$}}}\;
\raisebox{2pt}{\mbox{\LARGE $\nwarrow$}}
&{}&\mbox{\LARGE $\nearrow$}&{}\\
{}&{}&\kern-10pt{\rm Virasoro}\kern-10pt&{}&{}
\EA\label{sl2diagr}
\EE
where the $\N2$ superconformal algebra is realized precisely by adding
to the Virasoro algebra the free fields of the non-critical string in
the conformal gauge.  On the other hand, reconstructing the affine
$\SL2$ currents takes yet another free boson~\cite{[S-sing]} (the
downward arrow being the Hamiltonian reduction).  The main variation that
we have in the $\N2$ string, apart from the $(\N0)\to(\N2)$
`translation', is that the $\SSL21$ and $\N4$ algebras are realized on
the {\it same\/} space of fields, which means that the upward arrows
denote `dressing' of the $\N2$ algebra with the same collection of
free fields in both cases, namely with the Liouville and ghost
multiplets of the $\N2$ string:
\BE\BA{rcccl}
\SSL21\kern-10pt&{}&{}&{}&\kern-10pt\N4\\
{}&\rlap{\raisebox{-2pt}{\mbox{\LARGE $\searrow$}}}\;
\raisebox{2pt}{\mbox{\LARGE $\nwarrow$}}
&{}&\mbox{\LARGE $\nearrow$}&{}\\
{}&{}&\kern-10pt\N2\ \mbox{`matter'}\kern-10pt&{}&{}
\EA\label{sl21diagr}
\EE

It should be recalled that in \req{sl2diagr}, a relation between, in
that case, the affine $\SL2$ and $\N2$ singular vectors exists, but is
far from trivial: a subclass of $\N2$ singular vectors are isomorphic
to the $\SL2$ singular vectors \cite{[S-sing],[FST]}, while the `bulk'
of $\N2$ singular vectors are related to singular vectors in $\SL2$
modules that are not of the usual highest-weight type, but rather have
infinitely many equivalent `almost-highest-weight' vectors.  The $\N2$
counterpart of these states are the {\it extremal\/} vectors~\cite{[FS]}.

\smallskip

On the other hand, in the diagram~\req{sl21diagr}, where the $\N2$
algebra is a `primitive' ingredient, little is known,
beyond~\cite{[BLLS]}, about its $\N4$ side\,\footnote{An intriguing
point, however, is that the construction of $\SSL21$ singular vectors
that we will consider below is also similar to a construction of
certain $\N4$ singular vectors, even though the latter has been
realized only in a very special case~\cite{[ST1]}.}. As to the
$\SSL21$ algebra, we will show that it combines, in a rather
non-trivial way, certain properties of the $\SL2$ and $\N2$
algebras. In particular, the extremal states of the $\N2$ algebra lift
naturally to the larger algebras. This leads to very suggestive
similarities between the theories of $\SSL21$ and $\N2$ singular
vectors, both being closely related to the respective extremal
vectors.  The analysis of extremal vectors does immediately produce
certain series of singular vectors, by a kind of
`multiplet-shortening' mechanism, and is actually very suggestive as
to how the general construction for singular vectors can be built. The
latter takes introducing the `continued' extremal vectors and,
accordingly, the continued operators that generate them from the
vacuum. These operators are realized in terms of `continued products'
of fermions.

\smallskip

Thus the idea to consider extremal vectors~\cite{[FS]}, elaborated in
application to the $\N2$ superconformal algebra in collaboration with
I.~Tipunin~\cite{[ST3]}, turns out to be very efficient for the affine
$\SSL21$ as well (and looks quite promising also for any algebra with
at least two fermionic currents, in fact with a spectral flow
transform).  It may be hoped that this analysis will be useful to
relate the $\SSL21$ and $\N2$ fusion rules (for the related material,
see~\cite{[AY],[Andreev],[Petersen],[Gd]}).

\smallskip

The extremal states can be viewed as a generalization of different
{\it pictures\/} \cite{[FMS]} to the case of non-free fermions. Recall
that for the free first-order bosonic systems, different pictures are
inequivalent in the sense of Verma modules, while for free fermions,
on the contrary, they {\it are\/} equivalent. For the interacting
fermions, the situation turns out to be somewhat `intermediate':
generically, the extremal states are still equivalent to each other,
but it may (and does) happen for some values of the relevant
parameters (the weights) that the equivalence breaks down. To continue
with the analogy with the first-order free bosons and fermions, recall
that the pictures in the bosonic case are changed by the exponential
of a current that participates in `bosonizing' the system,
$\exp\phi$. By considering operators like $\exp\alpha\phi$, we can
change the picture arbitrarily (at least in principle, at the expense
of non-localities). The same is true for bosonized free fermionic
system. However, when the fermions are non-free, such a bosonization no longer
exists, and changing the picture by an arbitrary number
is realized by the `continued' products of modes of the fermionic
generators. For the $\N2$ superconformal algebra,
for example, we have two fermionic currents $\cQ_m$ and $\cG_n$,
with
$$
\{\cG_m,\,\cQ_n\}=2\cL_{m+n}-2n\cH_{m+n}+
\frac{\ctop}{3}(m^2+m)\delta_{m+n,0}\,,
$$
and the `continued' operators $q(a,b)$ and $g(a,b)$ can heuristically
be thought of as $\prod_a^b\,\cQ_\mu$ and $\prod_a^b\,\cG_\mu$
respectively. Had $\cQ$ and $\cG$ been free and bosonized through a
free scalar, these `continued' operators would have been constructed
`explicitly'. When no such bosonization is possible, however, the set
of bosonization rules can nevertheless be replaced by a set of
{\it algebraic rules\/} to deal with the new operators. For instance, when
considering the commutator of the Virasoro generators $\cL_n$ with
$g(a,b)$, we can first take $a$ and $b$ such that $b-a=N$ is a
positive integer, then we observe that in the commutator
$$
[\cL_1,\, \cG_a\,\ldots\,\cG_{a+N}]
$$
only the commutator $[\cL_1, \cG_{a+N}]$ gives a non-zero
contribution; for $[\cL_2, \cG_a\,\ldots\,\cG_{a+N}]$, similarly, only
$\cG_{a+N-1}$ and $\cG_{a+N}$ would contribute non-vanishing results,
and so forth. This therefore extends to {\it arbitrary complex\/} $a$
and $b$ as~\cite{[ST2]}
$$
[\cL_n,\,g(a,b)]=\sum_{i=0}^{n-1} g(a,b-i-1)\,[\cL_n,\,\cG_{b-i}]\,
\cG_{b-i+1}\,\ldots\,\cG_b\,,\quad n\geq1
$$
(for negative $n$, one counts from the left-hand end). 

These rules can be developed into a consistent algebraic scheme, which
does in a sense play the role of bosonization of non-free fermions. It
has appeared in~\cite{[ST2]}, while the states that these operators
create from the vacuum have been introduced, and their
representation-theoretic importance was stressed, in~\cite{[FS]}; we
take over from~\cite{[FS]} the name {\it extremal\/} for these states.
Let us also mention that the role of extremal states is
absolutely essential already in more `classical' problems, like that of
characterizing the image of the affine $\SL2$ \hw{} modules under the
Kazama--Suzuki mapping~\cite{[FST]}.

\bigskip

This paper is organized as follows.  In Section~2, we consider
highest-weight and extremal vectors in the affine $\SSL21$ Verma
modules. In Section~3, we present the general construction of the
$\SSL21$ singular vectors. As we have remarked, the singular vectors
are essentially divided onto those read off directly from the extremal
vectors, and those which require a `continuation'. The continued part
of the construction can be given in a direct analogy with the MFF
recipe for bosonic algebras~\cite{[MFF]}, however a nice feature of
$\SSL21$ as a superalgebra is the role of the fermionic currents in
the construction (in particular, the MFF-like continuation can be
performed entirely in terms of `continued' products of the
fermions). Section~4 is devoted to the analysis of representations of
$\SSL21$ realized on the noncritical $\N2$ string worldsheet by
tensoring the $\N2$ superconformal matter with the corresponding free
fields. An essential point here is that no assumptions are made on the
nature of the `matter' $\N2$ superconformal theory, in particular its
currents are not supposed to be `bosonized' in terms of free fields;
in that sense, the realization of $\SSL21$ we are going to consider is
{\it not\/} a free-field realization.  This will be important in
Section~5, where we consider how the $\SSL21$ singular vectors from
Section~3 evaluate in this realization. Again in contrast with
free-field realizations, none of the $\SSL21$ singular vectors
vanishes; in fact, they reduce precisely to singular vectors of the
underlying $\N2$ superconformal algebra, except for a particular
series labelled by positive integers.  

\section{The affine $\SSL21$ algebra}
\subsection{$\SSL21$ commutators,  automorphisms, and the spectral flow}\lvm
The algebra consists of four bosonic currents, $E^{12}$, $\Hminus$,
$F^{12}$, $\Hplus$, and four fermionic ones, $E^1$, $E^2$, $F^1$, and
$F^2$.  We will quite often drop `affine' when referring to this
algebra.  The non-vanishing commutation relations (with the brackets
$[~,~]$ denoting the {\it super\/}commutator) read
\BE\new\BA{rclrcl}\!

{[}\Hminus_m, E^{12}_n] &=& E^{12}_{m+n}\,,&
{[}\Hminus_m, F^{12}_n] &=& -F^{12}_{m+n}\,,\\
{[}E^{12}_m, F^{12}_n] &=& m \delta_{m+n, 0} k + 2 \Hminus_{m+n}\,,&
{[}H^\pm_m, H^\pm_n] &=& \mp\half m \delta_{m+n, 0} k\,,\\
{[}F^{12}_m, E^2_n] &=& F^1_{m+n}\,,&
{[}E^{12}_m, F^2_n] &=& -E^1_{m+n}\,,\\
{[}F^{12}_m, E^1_n] &=& -F^2_{m+n}\,,&
{[}E^{12}_m, F^1_n] &=& E^2_{m+n}\,,\\

{[}H^\pm_m, E^1_n] &=& \half  E^1_{m+n}\,,&
{[}H^\pm_m, F^1_n] &=& -\half  F^1_{m+n}\,,\\

{[}H^\pm_m, E^2_n] &=& \mp\half  E^2_{m+n}\,,&
{[}H^\pm_m, F^2_n] &=& \pm\half  F^2_{m+n}\,,\\
{[}E^1_m, F^1_n] &=& \multicolumn{4}{l}{-m \delta_{m+n, 0} k + \Hplus_{m+n} -
                        \Hminus_{m+n}\,,}\\
{[}E^2_m, F^2_n] &=& \multicolumn{4}{l}{m \delta_{m+n, 0} k + \Hplus_{m+n} +
                        \Hminus_{m+n}\,,}\\
{[}E^1_m, E^2_n] &=& E^{12}_{m+n}\,,&
{[}F^1_m, F^2_n] &=& F^{12}_{m+n}\\
\EA\EE
The affine $\SL2$ subalgebra is thus generated by $E^{12}_m$,
$\Hminus_m$, and $F^{12}_m$, and it commutes with the $U(1)$
subalgebra generated by $\Hplus_m$.

The affine $\SSL21$ algebra admits the following order-four automorphism
\BE\new\BA{rclrclrcl}
E^1&\mapsto&-F^1\,,&E^2&\mapsto&-F^2\,,&E^{12}&\mapsto&F^{12}\,,\\
F^1&\mapsto&E^1\,,&F^2&\mapsto&E^2\,,&F^{12}&\mapsto&E^{12}\,,\\
\Hminus&\mapsto&-\Hminus\,,&\Hplus&\mapsto&-\Hplus\,,
\EA\label{4thorder}\EE
and also an involutive automorphism:
\BE\new\BA{rclcrcl}
E^1_n&\mapsto& E^2_n\,,&& E^2_n&\mapsto& E^1_n\,,\\
F^1_n&\mapsto& -F^2_n\,,&& F^2_n&\mapsto& -F^1_n\,,\EA
\quad \Hplus_n\mapsto-\Hplus_n
\label{auto}\EE
(the other generators remain unchanged). 

Further, the spectral flow transform
\BE
\cU_\theta:{}\new\BA{rclcrcl}
E^1_n&\mapsto& E^1_{n-\theta}\,,&& E^2_n&\mapsto& E^2_{n+\theta}\,,\\
F^1_n&\mapsto& F^1_{n+\theta}\,,&& F^2_n&\mapsto& F^2_{n-\theta}\,,\EA\quad
\Hplus_n\mapsto\Hplus_n + k\theta\delta_{n,0}
\label{spectral}\EE
(where, again, the $\SL2$ subalgebra is inert) produces an isomorphic
algebra for any $\theta\in\oC$.  In any of the thus obtained
isomorphic algebras, the fermions have integer-spaced modes with the
`offset' $\pm\theta$.  In particular, for $\theta=\half$, we will have
the `Ramond' algebra.  For $\theta\in\oZ$, the spectral flow transform
is an automorphism of the algebra.  As we will see, the transformation
\req{spectral} is consistent with the $\N2$ superconformal spectral
flow transform~\cite{[SS],[LVW]}.

\medskip

Let us note finally that in terms of the normal-ordered field operators
$C(z)=\sum_{n\in\oZ}C_nz^{-n-1}$, where $C=(E^1,E^2,
E^{12},\Hminus,\Hplus,F^1,F^2,F^{12})$, the Sugawara \emt{} reads
\begin{eqnarray}
T_{\rm Sug} &=& \frac{1}{k+1}\Bigl(\Hminus\, \Hminus - \Hplus\,\Hplus +
\half E^{12}\, F^{12} + \half F^{12}\, E^{12} +
\half E^1\,F^1 - \half F^1\,E^1 -
\half E^2\,F^2 + \half F^2\,E^2\Bigr)\nonumber\\
{}&=&{}
\frac{1}{k+1}\Bigl(\Hminus\, \Hminus - \Hplus \Hplus
+ E^{12}\, F^{12} + E^1\, F^1  - E^2\, F^2  \Bigr)
\label{N2Sug}
\end{eqnarray}

\subsection{$\SSL21$ \hw{} modules and extremal states}\lvm
Consider the \hw{} conditions. They have to be imposed in such a way as not to
overdetermine the system of constraints. As usual, this selects essentially
the positive-moded generators, with some subtleties arising with the lowest
annihilators (in some formulations, these will be viewed as `zero modes'
of the vacuum). As to the generators $\Hplus$ and $\Hminus$, the
standard Heisenberg module \hw{} conditions read
\BE
\Hplus_{\geq1}\approx0\,,\qquad\Hminus_{\geq1}\approx0\,.
\label{hwH}
\EE
A characteristic feature of fermionic systems, on the other hand, is
that the distinction between creation and annihilation operators can
be drawn arbitrarily, recall for instance the `$q$'-vacua of the $bc$
systems \cite{[FMS]}.  Although the above $E^1, E^2, F^1$ and $F^2$
are not {\it free\/} fermions, a similar effect does take place
here as well (see the Introduction). For an arbitrary $\theta$, we can
choose a vacuum on which $E^1_{\geq-\theta+{1\over2}}$ and
$F^1_{\geq\theta+{1\over2}}$ would be annihilators\,\footnote{Note
that Eqs.~\req{auto} allow us to change the roles of the two pairs
(${}^1$ and ${}^2$) of the fermionic generators.}. Then, the strongest
conditions we can have for $E^2$ and $F^2$ would be
$E^2_{\geq\theta-{1\over2}}\approx0$ and
$F^2_{\geq-\theta+{3\over2}}\approx0$.

Note that the fermionic annihilation conditions imply those for the bosons as
\BE\new\BA{rclrclcrcl}
E^1_{\geq-\theta+{1\over2}}&\approx&0\,,&E^2_{\geq\theta-{1\over2}}&\approx&0\,,
&\Longrightarrow&E^{12}_{\geq0}&\approx&0\,,\\
F^1_{\geq\theta+{1\over2}}&\approx&0\,,&F^2_{\geq-\theta+{3\over2}}&\approx&0\,,
&\Longrightarrow&F^{12}_{\geq2}&\approx&0\,.
\EA\label{hwraw}\EE
However, the resulting \hw{} condition in the $\SL2$ subalgebra are
{\it not\/} those of the standard Verma modules; on the contrary, the
$\SL2$ modules with a `\hw' state determined by the right
column in \req{hwraw} do in fact have infinitely many equivalent
highest-weight vectors, and thus should not be called Verma modules,
once the latter are understood to have a unique \hw{} vector.

The \hw{} conditions \req{hwraw} (together with \req{hwH}) can be
strengthened consistently, so as to yield the standard Verma modules for
the $\SL2$ subalgebra, as
\BE\new\BA{rclrclrcl}
E^1_{\geq-\theta+{1\over2}}\ket{p,j,k;\theta}&=&0\,,&
E^2_{\geq\theta-{1\over2}}\ket{p,j,k;\theta}&=&0\,,
&E^{12}_{\geq0}\ket{p,j,k;\theta}&=&0\,,\\
F^1_{\geq\theta+{1\over2}}\ket{p,j,k;\theta}&=&0\,,&
F^2_{\geq-\theta+{3\over2}}\ket{p,j,k;\theta}&=&0\,,
&F^{12}_{\geq1}\ket{p,j,k;\theta}&=&0\,,\\
\multicolumn{9}{c}{\Hplus_0\ket{p,j,k;\theta}~{}={}~(p-k\theta)\,
\ket{p,j,k;\theta}\,,
\qquad\Hminus_0\ket{p,j,k;\theta}~{}={}~j\,\ket{p,j,k;\theta}\,.}
\EA\label{hwmassive}\EE
These will be called the generalized \hw{} conditions, termed
`generalized' for the presence of the $\theta$ parameter.  The
`ordinary' \hw{} vector will be denoted as
$\ket{p,j,k}\equiv\ket{p,j,k;0}$; as we are going to see, the
`standard' choice of $\theta=0$ is merely a convention:
\begin{thm}\label{unless}
Unless $-p\pm j +\frac{k}{2} - r(k+1)=0$, $r\in\oZ$, the \hw{} conditions
\req{hwmassive} are equivalent for all $\theta\in\oZ$.
\end{thm}
Indeed, applying the fermionic modes to a chosen \hw{} state, we can reach all
the other integer-spaced \hw{} states. Define
\BE
\ket{p + (k+1)r,j,k;r}{\,\tilde{}\,}=
  \left\{\new\!\!\BA{ll}
    E^2_{r - {1\over2}}\,\ldots\,E^2_{-{3\over2}}\cdot
     F^1_{r + {1\over2}}\,\ldots\,F^1_{-{1\over2}}
\ket{p,j,k}\,, & r \leq 0\,,\\
    E^1_{-r + {1\over2}}\,\ldots\,E^1_{-{1\over2}} \cdot
      F^2_{-r + {3\over2}}\,\ldots\,F^2_{{1\over2}}\,
\ket{p,j,k}\,, & r\geq1
\EA\right.
\label{gen}\EE
\begin{lemma}
The states \req{gen} satisfy the generalized \hw{} conditions
\req{hwmassive}.
\end{lemma}
We thus recover all the generalized \hw{} states as {\it extremal\/}
states.  All these extremal states are `connected' to $\ket{p,j,k}$,
and hence to each other, as described by the following simple lemma,
which will be of fundamental importance however:
\begin{lemma}
Given a state of the form \req{gen}, the original \hw{} state 
can be reconstructed as
\BE
\new\BA{rcll}
\multicolumn{4}{l}{E^1_{{1\over2}}\,\ldots\,E^1_{-r - {1\over2}}\,
 F^2_{{3\over2}}\,\ldots\,F^2_{-r + {1\over2}}
  \ket{p + (k+1)r,j,k;r}{\,\tilde{}\,}}\\
\qquad\qquad&=&
    \prod_{i=r}^{-1}
\left(i(k+1)+j+p-\frac{k}{2}\right)\left((i+1)(k+1)-j+p-\frac{k}{2}\right)
\,,&r\leq0
\\
\multicolumn{4}{l}{E^2_{-{1\over2}}\,\ldots\,E^2_{r - {3\over2}}\,
 F^1_{{1\over2}}\,\ldots\,F^1_{r - {1\over2}}\,
  \ket{p + (k+1)r,j,k;r}{\,\tilde{}\,}}\\
&=&
   \prod_{i=1}^r
\left((i-1)(k+1)+j+p-\frac{k}{2}\right)\left(i(k+1)-j+p-\frac{k}{2}\right)
\,,&r\geq1
\EA\label{travel}\EE
\end{lemma}
One can therefore travel along the integer-spaced generalized \hw{}
vectors {\it as long as none of the factors in the above formulae vanishes\/}:
\BE
\unitlength=1pt
\begin{picture}(250,40)
\put(-35,15){\Huge $\ldots$}
\put(0,0){$\times$}
\put(27,30){\vector(-1,-1){24}}
\put(8,5){\vector(1,1){24}}
\put(30,30){$\bullet$}
\put(62,8){\vector(-1,1){24}}
\put(36,28){\vector(1,-1){24}}
\put(60,0){\put(0,0){$\times$}
           \put(27,30){\vector(-1,-1){24}}
           \put(8,5){\vector(1,1){24}}
           \put(30,30){$\bullet$}
           \put(62,8){\vector(-1,1){24}}
           \put(36,28){\vector(1,-1){24}}
}
\put(140,42){$\ket{p,j,k}$}
\put(125,22){${}^{F^1_{\!\!-{1\over2}}}$}
\put(105,22){${}^{E^2_{\!\!-{3\over2}}}$}
\put(65,22){${}^{F^1_{\!\!-{3\over2}}}$}
\put(45,22){${}^{E^2_{\!\!-{5\over2}}}$}
\put(5,22){${}^{F^1_{\!\!-{5\over2}}}$}
\put(157,0){${}^{F^2_{\!\!{1\over2}}}$}
\put(198,0){${}^{E^1_{\!\!-{1\over2}}}$}
\put(217,0){${}^{F^2_{\!\!-{1\over2}}}$}
\put(258,0){${}^{E^1_{\!\!-{3\over2}}}$}
\put(277,0){${}^{F^2_{\!\!-{3\over2}}}$}
\put(120,0){\put(0,0){$\times$}
           \put(27,30){\vector(-1,-1){24}}
           \put(8,5){\vector(1,1){24}}
           \put(30,30){$\bullet$}
           \put(62,8){\vector(-1,1){24}}
           \put(36,28){\vector(1,-1){24}}
}
\put(180,0){\put(0,0){$\times$}
           \put(27,30){\vector(-1,-1){24}}
           \put(8,5){\vector(1,1){24}}
           \put(30,30){$\bullet$}
           \put(62,8){\vector(-1,1){24}}
           \put(36,28){\vector(1,-1){24}}
}
\put(240,0){\put(0,0){$\times$}
           \put(27,30){\vector(-1,-1){24}}
           \put(8,5){\vector(1,1){24}}
           \put(30,30){$\bullet$}
           \put(62,8){\vector(-1,1){24}}
           \put(36,28){\vector(1,-1){24}}
}
\put(310,15){\Huge$\ldots$}
\end{picture}
\label{traveldiagr}
\EE
The solid dots represent the states \req{gen}, labelled by
$\theta\in\oZ$, positive or negative depending on whether we move
right or left (while the crosses will be considered later, in
Section~\ref{subsec:mff}).  For some values of the parameters
involved, the `transitivity' among these states fails. This happens at
`points of no return', when the generator that does generically shift
to an adjacent extremal state turns out to be an
annihilator. Therefore, at the points where the transitivity breaks
down, additional highest-weight-type relations are satisfied; these do
overconstrain the system in the sense that they require certain
relations between the parameters $p$, $j$, and $k$ of the \hw{} state;
these relations thus guarantee the existence of a singular vector over
that \hw{} state, since the breakdown of transitivity means that no
operator in the representation of the algebra would map back to the
\hw{} state we started with; the resulting {\it subrepresentation\/}
is conveniently encoded by the respective singular
vector\,\footnote{To avoid misunderstanding, let us note once again
that {\it not all\/} singular vectors follow by such a straightforward
procedure, however, as we will see shortly.}.

In fact, as soon as one realizes that a \hw{} vector should be thought
of as a whole class of equivalent but not identical, states, one also
concludes that the same is obviously true for {\it singular\/}
vectors. Once we have encountered in the diagram \req{traveldiagr} a
point where the transitivity breaks down, there would be another such
{\it subdiagram\/} growing out of that point, and the respective singular
vector is in fact a class of states belonging to that subdiagram. In
the $\N2$ case, where things are easier to visualize, this will be
represented as in~\req{branchdiag} and~\req{degendiagram}.

\section{Singular vectors of the affine $\SSL21$ algebra at arbitrary 
level\label{sec:sl21sing}}\lvm We will now construct singular vectors
in the $\SSL21$ Verma modules. As is the case with the $\N2$
superconformal algebra, there are singular vectors which follow
immediately from the extremal vectors~\cite{[ST3]}, and we will take
over from the $\N2$ algebra the name `charged' for these; the
remaining $\SSL21$ singular vectors will be called the MFF ones, since
their construction is very similar to the standard MFF
formulae~\cite{[MFF]}.  In what follows, we will stress the role of
extremal vectors considered in the previous section.

\subsection{The `charged' $\SSL21$ singular vectors}\lvm
A part of the $\SSL21$ singular vectors can be immediately read off
from the diagram \req{traveldiagr}.  Obviously, the `points of no
return' mean that we have a {\it submodule\/} in the original Verma
module, and therefore such points are associated with a singular
vector.  From the above, we easily see that the positive-$r$ and
negative-$r$ cases match to produce the following `special' values of
the parameters:
\BE\new\BA{rcll}
\ptop_2(r,j,k)&=&-j + \half k - r(k+1)\,,&r\in\oZ\,,\\
\ptop_1( r,j,k)&=& j + \half k -  r\,(k+1)\,,& r\in\oZ\,,
\EA\label{ptop}\EE
at which singular vectors occur.

The singular vectors themselves are also read off from 
the construction \req{gen} of the extremal states:
\BE
\ketch2{E( r, j, k)}=\left\{\new\!\!\BA{ll}
\underbrace{E^2_{ r-{1\over2}}\,\ldots\,E^2_{-{3\over2}}}_{- r}\cdot
\underbrace{F^1_{ r+{1\over2}}\,\ldots\,F^1_{-{1\over2}}}_{- r}\,
\ket{\,\ptop_2( r,j,k),j,k},& r\leq-1\,,\\
\underbrace{E^1_{- r+{1\over2}}\,\ldots\,E^1_{-{1\over2}}}_{ r}
\cdot
\underbrace{F^2_{- r+{1\over2}}\,F^2_{- r+{3\over2}}\,\ldots\,
F^2_{{1\over2}}}_{ r+1}\,
\ket{\,\ptop_2( r,j,k),j,k},& r\geq0\,,
\EA\right.\quad r\in\oZ
\label{Ech2}\EE
These will be called the charged-II singular vectors.

\begin{lemma}
The charged-II singular vectors $\ketch2{E( r, j, k)}$
satisfy the $\theta=r$-case of the \hw{}
conditions
\BE\new\BA{rclrclrclrcl}
E^1_{\geq- \theta+{1\over2}}&\approx&0\,,&
E^2_{\geq \theta-{1\over2}}&\approx&0\,,
&E^{12}_{\geq0}&\approx&0\,,&\Hplus_{\geq1}&\approx&0\,,\\
F^1_{\geq \theta+{1\over2}}&\approx&0\,,&
F^2_{\geq- \theta+{1\over2}}&\approx&0\,,
&F^{12}_{\geq1}&\approx&0\,,&\Hminus_{\geq1}&\approx&0\,,
\EA\quad (\theta - \half)\,k + \Hminus_0 + \Hplus_0 \approx0
\label{hwtop2}\EE
\end{lemma}
These special \hw{} conditions, which are stronger than
\req{hwmassive}, will generally be referred to as {\it topological\/},
the terminology taken over from the case of the $\N2$ superconformal
algebra. As we are going to see, there exist two types of the
topological \hw{} conditions, and, accordingly, two types of the
associated singular vectors. More precisely, equations \req{hwtop2}
will be called the generalized {\it topological-II\/} \hw{}
conditions. The `ordinary' topological-II \hw{} conditions will, as
usual, be the $\theta=0$-case of the generalized ones. As explained
above, the appearance of the topological \hw{} conditions reflects the
existence of a relation between the parameters of the \hw{} state.

We thus see that the $r$th charged singular vector satisfies the
$\theta=r$-case of the generalized topological-II \hw{} conditions.
This breaks down the transitivity in \req{traveldiagr} and is therefore
sufficient for the corresponding state to be singular. One may also
wish to choose a system of representatives of singular vectors that
would satisfy a single, `fixed' rather than `relative', set of \hw{}
conditions. These \hw{} conditions cannot then be topological;
instead, it is rather natural to choose the $\theta=0$ case of the
\hw{} conditions~\req{hwmassive}.  We then in a standard way 
(see~\req{traveldiagr}), find the
$\theta=0$-representative, as
\BE \ket{S( r, j, k)}^{(2)}_{{\rm
ch}}=\left\{ \!\!\new\BA{ll} E^1_{{1\over2}}\,\ldots\,E^1_{- r -
{1\over2}}\, F^2_{{3\over2}}\,\ldots\,F^2_{- r - {1\over2}}\,\ket{E(
r, j, k)}^{(2)} \,,& r\leq-1\,,\\ E^2_{-{1\over2}}\,\ldots\,E^2_{ r -
{3\over2}}\, F^1_{{1\over2}}\,\ldots\,F^1_{ r - {1\over2}}\,\ket{E( r,
j, k)}^{(2)} \,,& r\geq0\,, \EA\right.\quad
r\in\oZ\,.
\label{ground2}\EE
As is easy to check, these do indeed satisfy the $\theta=0$-case of the \hw{}
conditions~\req{hwmassive}.

As we see from this example, the topological representative of a
singular vector is the `minimal' one in that all the other
representatives are obtained by making more steps along the the
diagram of the type of~\req{traveldiagr}, already {\it inside\/} the
{\it sub\/}module.

\medskip

Similarly, we define the charged-I singular vectors by
\BE
\ketch1{E( r, j, k)}=\left\{\new\!\!\BA{ll}
\underbrace{E^2_{ r - {1\over2}}\,\ldots\,E^2_{-{3\over2}}}_{- r}\cdot
\underbrace{F^1_{ r - {1\over2}}\,F^1_{ r + {1\over2}}\,\ldots\,
F^1_{-{1\over2}}}_{- r+1}\,
\ket{\,\ptop_1( r,j,k),j,k},& r\leq0\,,\\
\underbrace{E^1_{- r + {1\over2}}\,\ldots\,E^1_{-{1\over2}}}_{ r}\cdot
\underbrace{F^2_{- r + {3\over2}}\,\ldots\,F^2_{{1\over2}}}_{ r}\,
\ket{\,\ptop_1( r,j,k),j,k},& r\geq1\,,
\EA\right.\quad r\in\oZ
\label{Ech1}\EE
{}From Eqs.~\req{travel} we have
\begin{lemma}
The charged-I singular vectors~\req{Ech1} satisfy the \hw{}
conditions
\BE\new\BA{rclrclrclrcl}
E^1_{\geq- \theta+{1\over2}}&\approx&0\,,&
E^2_{\geq \theta-{1\over2}}&\approx&0\,,
&E^{12}_{\geq0}&\approx&0\,,&\Hplus_{\geq1}&\approx&0\,,\\
F^1_{\geq \theta-{1\over2}}&\approx&0\,,&
F^2_{\geq- \theta+{3\over2}}&\approx&0\,,
&F^{12}_{\geq1}&\approx&0\,,&\Hminus_{\geq1}&\approx&0\,,
\EA\quad ( \theta - \half)\,k - \Hminus_0 + \Hplus_0\approx0\,.
\label{hwtop1}\EE
\end{lemma}
These will be called the {\it topological-I\/} \hw{} conditions. As
before, the $\theta=0$-represent\-at\-ives of the charged-I
singular vectors follow by taking a further trip over the diagram of
the type of~\req{traveldiagr}, and read
\BE
\ket{S( r, j, k)}^{(1)}_{{\rm ch}}=\left\{
\!\!\new\BA{ll}
 E^1_{{1\over2}}\,\ldots\,E^1_{- r - {1\over2}}\,
   F^2_{{3\over2}}\,\ldots\,F^2_{- r + {1\over2}}\,\ket{E( r, j, k)}^{(1)}
\,,& r\leq-1\,,\\
 E^2_{-{1\over2}}\,\ldots\,E^2_{ r - {3\over2}}\,
   F^1_{{1\over2}}\,\ldots\,F^1_{ r - {3\over2}}\,\ket{E( r, j, k)}^{(1)}
\,,& r\geq0\,,
\EA\right.\qquad r\in\oZ\,.\label{ground1}\EE
(and satisfy the $\theta=0$-case of the \hw{} conditions~\req{hwmassive}).

It is readily seen that the singular vectors \req{hwtop2} and
\req{hwtop1} are related by a combination of the automorphism
\req{auto} and the spectral flow transform \req{spectral} with
$\theta=-1$; the same is obviously true for the vectors \req{ground2}
and~\req{ground1}:
\BE\new\BA{rcl}
\ket{S(r, j, k)}^{(2)}_{{\rm ch}}&\mapsto&\ket{S(-r, j, k)}^{(1)}_{{\rm ch}}\\
(E^1_n\,,\,E^2_n \,,\,F^1_n\,,\,F^2_n)&\mapsto&
(E^2_{n+1}\,,\,E^1_{n-1}\,,\,-F^2_{n-1}\,,\,-F^1_{n+1})\\
\Hplus_n&\mapsto&-\Hplus_n+k\delta_{n,0}
\EA\EE

The results of this subsection can be summarized as the following
theorem, which we formulate explicitly even though it is essentially a
reformulation of~\reb{unless}:
\begin{thm}\label{charged}
Whenever $p\pm j-\frac{k}{2}=-r(k+1)$, $r\in\oZ$, the $\SSL21$ Verma
module with the \hw{} vector $\ket{p,j,k}$ has a singular vector,
whose topological representative is given by~\req{Ech2} or \req{Ech1},
and the minimal-level representative, by~\req{ground2}
or~\req{ground1} respectively.
\end{thm}

\subsection{The $\SSL21$ MFF construction\label{subsec:mff}}\lvm
We now proceed to the `continued' part of the construction of $\SSL21$
singular vectors. Their positions (as well as positions of the charged
ones) are of course known from the Ka\v c--Kazhdan determinant
\cite{[KK]}, which for the $\SSL21$ algebra has recently been reviewed
in \cite{[BT]} (see also references therein), where an important step
was made by proposing an MFF-like construction for singular vectors in
the case of rational $k+1=\frac{p}{q}$, and classifying a number of
Verma module embedding diagrams. In what follows, we will proceed for
the general (complex) $k$ and will use the continued commutation
relations in order to formulate the general construction for the
$\SSL21$ singular vectors\footnote{As we have remarked, in the case of
rational $k$ one applies the MFF-like construction as many times as
the rationality of $k$ allows one to (i.e., representing the spin $j$
as the RHSs of \req{jpm} with different $r$ and $s$).}.  At the most
fundamental level, the general MFF construction for $\SSL21$ can be
formulated solely in terms of the `continued' products of fermionic
generators, similarly to the $\N2$ case \cite{[ST3]}\,\footnote{and in
fact in the spirit of the suggestion of ref.~\cite{[ST1]} for (a
particular representation of) the $\N4$ algebra.}. In this paper,
however, we will not develop this consistently `fermionic' approach,
and will comment on it only in a remark.

So as not to interrupt the presentation later on, we begin with a
simple technical observation concerning the crosses in the
diagram~\req{traveldiagr}:
\begin{lemma}\label{modified}
Unless $p - j - \half k=0$, any massive \hw{} state
$\MW{p,j,k;\theta}$ is equivalent to a state
$\MW{p-\half,j-\half,k}^{\rm mod\,(1)}$ for which the modified \hw{}
conditions 
\BE
\new\BA{rclrclrclrcl}
E^1_{\geq-(\theta-1)+{1\over2}}&\approx&0\,,&
E^2_{\geq(\theta-1)+{1\over2}}&\approx&0\,,
&E^{12}_{\geq0}&\approx&0\,,&\Hplus_{\geq1}&\approx&0\,,\\
F^1_{\geq(\theta-1)+{1\over2}}&\approx&0\,,&
F^2_{\geq-(\theta-1)+{1\over2}}&\approx&0\,,
&F^{12}_{\geq1}&\approx&0&\Hminus_{\geq1}&\approx&0
\EA\label{hwmod2}\EE
hold. Similarly, unless $p + j - \half k=0$, the state
$\MW{p,j,k;\theta}$ is equivalent to a state
$\MW{p+\half,j-\half,k}^{\rm mod\,(2)}$, which satisfies the \hw{}
conditions \req{hwmod2} with $\theta\to\theta+1$.
\end{lemma}
Indeed, we build the sate $\ket{~}^{\rm mod\,(1)}$ as
\BE
\MW{p-\half,j-\half,k;\theta}^{\!\rm mod\,(1)} = 
F^1_{\theta-{1\over2}}\,\MW{p,j,k;\theta},\;~
{\rm then}\;~
\MW{p,j,k;\theta} = \frac{1}{p - j - {1\over2}k}
E^1_{-\theta+{1\over2}}\,\MW{p-\half,j-\half, k;\theta}^{\!\rm mod\,(1)}.
\EE
If, however, $p - j - \half k$ does vanish, then $\MW{p-\half,j-\half,
k;\theta}^{\rm mod\,(1)}$ satisfies the topological-I \hw{} conditions
and hence the massive \hw{} state cannot be recovered.

We now proceed to the generalization of the MFF construction
to~$\SSL21$.

\subsubsection*{Generic case}\lvm
The generalization of the MFF construction that we will propose is
motivated by the following observations.
\begin{lemma}
The state
\begin{eqnarray}
\ket{p+\half,j-r-\half,k,r}''&=&
\underbrace{F^1_{{1\over2}}\,\ldots\,F^1_{r-{1\over2}}}_{r}\cdot
\underbrace{F^2_{-r+{1\over2}}\,\ldots\,F^2_{{1\over2}}}_{r+1}\,
\ket{p,j,k}\nonumber\\
\noalign{\noindent\mbox{rewrites as}}
{}&=&
(F^{12}_0)^{r}\,F^2_{{1\over2}}\,\ket{p,j,k}\,,\qquad\qquad\qquad r\in\oN_0\,.
\label{test2}
\end{eqnarray}
\end{lemma}
Then going back from $\ket{~}''$ to the \hw{} state $\ket{p,j,k}$ can
be achieved as
\BE
E^2_{-{1\over2}}\,\ldots\,E^2_{r-{1\over2}}\cdot
E^1_{-r+{1\over2}}\,\ldots\,E^1_{-{1\over2}}\,\ket{p+\half,j-r-\half,k,r}''=
r!\,(j+p-\frac{k}{2})\prod_{i=1}^r(2j - i) \,\ket{p,j,k}
\label{primeprimeback}\EE
if none of the factors on the right-hand side vanishes; if one of them
does, this means, by the argument we have expanded above, that we
encounter a singular vector at a certain point on the way.  An
important point is that the possibilities for this to happen are of two
sorts: either $j+p-\frac{k}{2}=0$, which signifies a topological point
according to \reb{modified}, or one of the the $2j-i$ factors
vanishes. We now concentrate on the basic case when this latter factor
vanishes for $i=r$ (otherwise, one should simply notice an earlier
appearance of a singular vector and consider the same formulae for
smaller $r$).  This singular vector is going to be one of the key
ingredients of the MFF construction. However, the presence of the
first factor on the RHS of \req{primeprimeback} leads to an important
subtlety, which will be discussed on page~\pageref{subtle}.

Similarly, acting first with the modes of $F^1$ and then, $F^2$,
as\,\footnote{Here and in what follows, $\oN=\{1,2,\ldots\}$ and
$\oN_0=\{0\}\cup\oN$.}
\BE\new\BA{rclll}
\ket{p-\half,j-r-\half,k,r}'&=&
\underbrace{F^2_{{3\over2}}\,\ldots\,F^2_{r+{1\over2}}}_{r}\cdot
\underbrace{F^1_{-r-{1\over2}}\,\ldots\,F^1_{-{1\over2}}}_{r+1}\,
\ket{p,j,k}\,,
\\
{}&=&
(F^{12}_0)^{r}\,F^1_{-{1\over2}}\,\ket{p,j,k}\,,&{}&r\in\oN\,,
\EA\label{test1}\EE
we will have
\BE
E^1_{{1\over2}}\,\ldots\,E^1_{r+{1\over2}}\cdot
E^2_{-r-{1\over2}}\,\ldots\,E^2_{-{3\over2}}\,\ket{p-\half,j-r-\half,k,r}'=
r!\,(-j+p-\frac{k}{2})\prod_{i=1}^{r}(2j - i) \,\ket{p,j,k}\,,
\EE

The singular vector that can be read off from \req{test2} or
\req{test1} would not yet be of the standard type, instead it would
satisfy the {\it modified\/} \hw{} conditions~\req{hwmod2}.  To obtain
a singular vector that would satisfy the standard \hw{} conditions
\req{hwmassive} with $\theta=0$, we have to further act with the
appropriate modes of $E^1$ or $E^2$, which would give
\BE\new\BA{rcl}
E^1_{{1\over2}}\,\ket{p-\half,j-r-\half,k,r}'&=&
\left(r(F^{12}_0)^{r-1}\,F^2_{{1\over2}}\,F^1_{-{1\over2}}+
(p-j-\frac{k}{2})(F^{12}_0)^r\right)\,\ket{p,j,k}\\
E^2_{-{1\over2}}\,\ket{p+\half,j-r-\half,k,r}''&=&
\left(r(F^{12}_0)^{r-1}\,F^2_{{1\over2}}\,F^1_{-{1\over2}}+
(p+j-r-\frac{k}{2})(F^{12}_0)^r\right)\,\ket{p,j,k}
\EA\label{both}\EE
\begin{lemma} For $r=2j$, 
both expressions \req{both} become equal to
\BE
\left(2j(F^{12}_0)^{2j-1}\,F^2_{{1\over2}}\,F^1_{-{1\over2}}+
(p-j-\frac{k}{2})(F^{12}_0)^{2j}\right)\,\ket{p,j,k}\,,
\label{minusj}\EE
which satisfies the \hw{} conditions~\req{hwmassive}.
\end{lemma}
In the spirit of \req{traveldiagr}, we now have the following
`kite' diagram (the $r=2$ example):
\BE
\unitlength=1pt
\begin{picture}(140,105)
\put(60,100){
 \put(00,00){$\bullet$}
 \put(-1,-1){\vector(-1,-1){14}}
}
\put(36,90){${}^{F^1_{\!\!-{1\over2}}}$}
\put(40,80){
 \put(00,00){$\times$}
 \put(-1,-1){\vector(-1,-1){14}}
}
\put(16,70){${}^{F^1_{\!\!-{3\over2}}}$}
\put(20,60){
 \put(00,00){$\circ$}
 \put(-1,-1){\vector(-1,-1){14}}
}
\put(-6,50){${}^{F^1_{\!\!-{5\over2}}}$}
\put(60,100){
 \put(6,-1){\vector(1,-1){14}}
}
\put(72,90){${}^{F^2_{{1\over2}}}$}
\put(80,80){
 \put(00,00){$\times$}
 \put(6,-1){\vector(1,-1){14}}
}
\put(92,70){${}^{F^2_{-{1\over2}}}$}
\put(100,60){
 \put(00,00){$\circ$}
 \put(6,-1){\vector(1,-1){14}}
}
\put(112,50){${}^{F^2_{-{3\over2}}}$}
\put(120,40){
 \put(00,00){$\circ$}
 \put(-1,-1){\vector(-1,-1){14}}
}
\put(115,18){${}^{F^1_{{1\over2}}}$}
\put(100,20){
 \put(00,00){$\circ$}
 \put(-1,-1){\vector(-1,-1){14}}
}
\put(95,-2){${}^{F^1_{{3\over2}}}$}
\put(00,40){
 \put(00,00){$\circ$}
 \put(6,-1){\vector(1,-1){14}}
}
\put(-2,19){${}^{F^2_{{3\over2}}}$}
\put(20,20){
 \put(00,00){$\circ$}
 \put(6,-1){\vector(1,-1){14}}
}
\put(19,-1){${}^{F^2_{{5\over2}}}$}
\put(40,00){$\circ$}
\put(46,6){\vector(1,1){14}}
\put(48,-5){${}^{E^1_{{1\over2}}}$}
\put(80,00){$\circ$}
\put(60,20){$\bullet$}
\put(78,6){\vector(-1,1){14}}
\put(62,-5){${}^{E^2_{\!\!-{1\over2}}}$}
\end{picture}
\EE

The construction extending \req{test1} and \req{test2} to $r\leq-1$
reads
\BE\new\BA{rcl}
\ket{p,j+r,k,r}'&=&
\underbrace{E^2_{{1\over2}}\,\ldots\,E^2_{-r-{1\over2}}}_{-r}\,
\underbrace{E^1_{r+{1\over2}}\,\ldots\,E^1_{-{1\over2}}}_{-r}\,\ket{p,j,k}\\
{}&=&(E^{12}_{-1})^{-r}\,\ket{p,j,k}
\EA\qquad r=-1,-2\ldots
\label{testnegative}\EE
This satisfies the \hw{} conditions \req{hwmassive}, and hence produces a
singular vector, iff
\BE
r+2 j - k -1 = 0\,.
\label{k+1}\EE

Now the combination of the two ways to produce
new \hw{} states out of a given one, Eqs.~\req{minusj} and \req{testnegative},
gives the following patterns:
\BE\BA{rcl}
{}&{}&\ket{p,-j,k}\mapsto\ldots\mapsto\ket{p,-n(k+1)-j,k}\mapsto
\ket{p,(n+1)(k+1)+j,k}\mapsto\ldots\,,\quad n\geq0\\
{}&\nearrow&{}\\
\ket{p,j,k}&{}&{}\\
{}&\searrow&{}\\
{}&{}&\ket{p,k+1-j,k}\mapsto\ldots\mapsto\ket{p,n(k+1)-j,k}\mapsto
\ket{p,-n(k+1)+j,k}\mapsto\ldots\,,\quad n\geq1
\EA\label{proto}\EE

As in the bosonic MFF construction \cite{[MFF]}, we now drop the
condition that the corresponding $r$ be an integer at each step in the
above formulae, and require instead that only the whole sequence of
mappings lead to an element of the Verma module built on
$\ket{p,j,k}$. Then, the resulting eigenvalue of $\Hminus_0$ must
differ from $j$ by an integer (if the vector is bosonic) or a
half-integer (if the vector is fermionic); from \req{minusj} and
\req{testnegative} we see that the former case applies, and thus the
necessary conditions for the respective lines in \req{proto} to yield
a singular vector in the Verma module after a finite number of steps,
are
\BE\new\BA{rclcrclll}
-n(k+1)-2j&=&m\,,&\rm or&(n+1)(k+1)&=&m\,,&n\geq0\,,&m\in\oZ\,,\\
n(k+1)-2 j&=&m\,,&\rm or&-n(k+1)&=&m\,,&n\geq1\,,&m\in\oZ\,.
\EA\label{jprelim}\EE
The tentative singular vectors are then constructed as
the MFF monomials
\BE
\ldots\,
E^1_{{1\over2}}(F^{12}_0)^{2(k+1)+2j}F^1_{-{1\over2}}\cdot
(E^{12}_{-1})^{k+1+2j}\cdot
E^1_{{1\over2}}(F^{12}_0)^{2j}F^1_{-{1\over2}}\,
\ket{p,j,k}
\label{mffplus}\EE
and
\BE
\ldots\,(E^{12}_{-1})^{3(k+1)-2j}\cdot
E^1_{{1\over2}}(F^{12}_0)^{2(k+1)-2j}F^1_{-{1\over2}}\cdot
(E^{12}_{-1})^{k+1-2j}\,
\ket{p,j,k}
\label{mffminus}\EE
respectively,
with $j$ determined by \req{jprelim}. 
Further, we require that the eigenvalue of $\Hminus_0$ on
these states differ from $j$ by a {\it positive\/} integer for
\req{mffminus} (in analogy with \req{testnegative})
and a {\it negative\/} one for \req{mffplus}
(in analogy with \req{minusj}), which would also guarantee that
the factor in the middle of the MFF monomial would always 
be raised to a {\it positive\/} integer power.
Thus the spins in \req{mffplus}
and \req{mffminus} are, respectively,
\BE\new\BA{rcl}
\jplus(r,s,k)&=&\frac{r}{2}-\frac{s-1}{2}(k+1)\,,\\
\jminus(r,s,k)&=&-\frac{r}{2}+\frac{s}{2}(k+1)\,,
\EA\quad r,s\in\oN\label{jpm}\EE
Thus the singular vectors become
\BE\!\!\new\BA{rcl}
\ket{{\rm MFF}(r,s,p,k)}^+\kern-8pt&=&\kern-8pt
E^1_{{1\over2}}(F^{12}_0)^{r+(s-1)(k+1)}F^1_{-{1\over2}}\cdot
(E^{12}_{-1})^{r+(s-2)(k+1)}\cdot\ldots\cdot
E^1_{{1\over2}}(F^{12}_0)^{r-(s-1)(k+1)}F^1_{-{1\over2}}\cdot{}\\
{}&{}&\hfill{}\cdot\ket{p,\frac{r}{2}-\frac{s-1}{2}(k+1),k}\\
\ket{{\rm MFF}(r,s,p,k)}^-\kern-8pt&=&\kern-8pt
(E^{12}_{-1})^{r+(s-1)(k+1)}\cdot
E^1_{{1\over2}}(F^{12}_0)^{r+(s-2)(k+1)}F^1_{-{1\over2}}\cdot\ldots\cdot
(E^{12}_{-1})^{r-(s-1)(k+1)}\,\cdot{}\\
{}&{}&\hfill{}\cdot\ket{p,\frac{-r}{2}+\frac{s}{2}(k+1),k}\\
{}&{}&r,s\in\oN\,,\quad p\in\oC
\EA\label{mff}\EE

In order to give meaning to this algebraically continued construction,
we need several commutation relations:
\begin{lemma}\label{resolving}
For $n\in\oN$, the following commutation relations hold in
the universal enveloping algebra of $\SSL21$:
\BE\!\!\!\!\new\BA{rcl}
(F^{12}_0)^n\,E^{12}_m &=&
 \Bigl(-n (n - 1) F^{12}_m -
  2 n \Hminus_m\,F^{12}_0 + E^{12}_m\, 
F^{12}_0\,F^{12}_0\Bigr)(F^{12}_0)^{n-2}
\,,\\
(F^{12}_0)^n\,E^1_m &=& \Bigl(-n F^2_m + E^1_m\,F^{12}_0\Bigr)
(F^{12}_0)^{n-1}\,,\\
(F^{12}_0)^{n}\,E^2_m &=& \Bigl(n F^1_m + E^2_m\,F^{12}_0\Bigr)
(F^{12}_0)^{n-1}\,,\\
(F^{12}_0)^{n}\,\Hminus_m &=& \Bigl(n F^{12}_m + \Hminus_m\,F^{12}_0\Bigr)
(F^{12}_0)^{n-1}\,.
\EA\label{commutef}\EE
\end{lemma}

Similarly,

\begin{lemma}\label{resolving2}
For $n\in\oN$, the following commutation relations hold:
\BE\!\!\!\!\new\BA{rcl}
(E^{12}_{-1})^n\,F^{12}_m &=&
 \Bigl(-n (n - 1) E^{12}_{m-2} -
  k\,n\,\delta_{m - 1, 0} E^{12}_{-1} +
  2 n \Hminus_{m-1}\,E^{12}_{-1} + F^{12}_m\, E^{12}_{-1}\,E^{12}_{-1}\Bigr)
(E^{12}_{-1})^{n-2}\,,\\
(E^{12}_{-1})^n\,F^1_m &=& \Bigl(n E^2_{m-1} + F^1_m\,E^{12}_{-1}\Bigr)
(E^{12}_{-1})^{n-1}\,,\\
(E^{12}_{-1})^{n}\,F^2_m &=& \Bigl(-n E^1_{m-1} + F^2_m\,E^{12}_{-1}\Bigr)
(E^{12}_{-1})^{n-1}\,,\\
(E^{12}_{-1})^{n}\,\Hminus_m &=& \Bigl(-n E^{12}_{m-1} +
\Hminus_m\,E^{12}_{-1}\Bigr)(E^{12}_{-1})^{n-1}\,.
\EA\label{commutee}\EE
\end{lemma}
These relations are now to be extended to $n\in\oC$, by first
continuing to $n\in\oZ$ and then simply postulating them to hold for
arbitrary $n$.

While the steps leading to \req{mff} were merely a motivation, now
that the formulae \req{mff} are written down, we have the following
\begin{thm}\label{mffthm}
For generic (complex) values of the $U(1)$ charge $p$ the level~$k$,
and $r,s\in\oN$, the monomial expressions~\req{mff} determine singular
vectors in the Verma modules with the highest-weight vectors
$\ket{p,\frac{r}{2}-\frac{s-1}{2}(k+1),k}$ and
$\ket{p,\frac{-r}{2}+\frac{s}{2}(k+1),k}$ respectively. For $s\geq2$,
the non-integral powers of $E^{12}_{-1}$ and $F^{12}_0$ are `resolved'
using the formulae of Lemmas~\ref{resolving} and~\ref{resolving2}: the
repeated use of relations \req{commutee} and \req{commutef} in the
expressions for $\ket{{\rm MFF}(r,s,p,k)}^\pm$ leads eventually to the
standard `Verma' form of these vectors.  {\rm (No rearrangements are
needed for $s=1$; `generic' here refers to all values of $p$ and $k$
except the $s^2-s$ points \req{exceptional}, to be considered
separately.)}
\end{thm}

\begin{rem}
A common feature of the (continued) {\it monomial\/} constructions for
singular vectors~\cite{[MFF],[ST2],[ST3]} is that the \hw{} conditions
are formally fulfilled after the application of each subsequent
`continued' factor, starting with the first one acting on the \hw{}
state (in~\req{mff}, these are separated with dots). However, it is
only after the application of precisely as many factors as prescribed by
the entire formula that the resulting vectors would belong to the
Verma module.\label{monomialhw}
\end{rem}

\begin{rem}\label{fermionic}
It is instructive (although not very useful for practical
calculations) to rewrite \req{mff} in a purely fermionic form, using
the continued products of the fermionic generators:
\BE\kern-6pt\new\BA{rcl}
\ket{{\rm MFF}(r,s,p,k)}^+\kern-8pt&=&\kern-8pt
E^1_{{1\over2}}f^2(\frac{3}{2},r+\half+(s-1)(k+1))
 f^1(-r-\half-(s-1)(k+1),\half)\\
{}&{}&\quad{}\cdot\ldots\\
{}&{}&\qquad{}\cdot e^2(\half,-r-\half+(s-2)(k+1))e^1(r+\half-(s-2)(k+1),\half)\\
{}&{}&\qquad\quad\cdot
f^2(\frac{3}{2},r+\half-(s-1)(k+1))f^1((s-1)(k+1)-r-\half,-\half)
\ket{p,\jplus(r,s,k),k}\\
\EA\EE
and similarly for ${\rm MFF}^-$. Here, e.g.,
$f^1(a,b)\,\mbox{``{}}{=}\mbox{{}''}\,\prod_a^b F^1_\alpha$, which can
be given an algebraic meaning similarly to how this has been done for
the `continued products' of the $\N2$ fermionic
generators~\cite{[ST3]}, as discussed in the Introduction. Such
operators produce the `continued' extremal states out of the chosen
vacuum.  Note that the grouping of the factors in the last formula is
slightly different from the one that was implied when
constructing~\req{mff}: namely, the factor
$E^1_{{1\over2}}(F^{12}_0)^{\mu}F^1_{-{1\over2}}$ lends the left
$E^1_{{1\over2}}$ to the $E$-factors further on the left in the
formula.
\end{rem}

Singular vectors in the `Ramond' sector -- which we will use in the
next section -- or in any other `sector' follow by applying to the
above expressions the spectral flow transform; this can be done
directly in the monomial forms~\req{mff}, and \req{Ech2}
and~\req{Ech1}.

\begin{rem}\label{r0}
The $\ket{{\rm MFF}(r,s)}^\pm$ singular vectors can formally be
defined also for $r=0$; in that case, however, one readily finds that
they are proportional to the \hw{} state, for example
\BE\new\BA{rcl}
\ket{{\rm MFF}(0,s, p, k)}^+ &=& a(s, p, k) \ket{p, \jplus(0, s, k), k},\,,\\
a(s, p, k)&=&\left\{\new\BA{ll}
(p - \frac{k}{2})\prod_{i=1}^{(s - 1)/2}(p - \frac{k}{2} + i(k+1)) 
(p - \frac{k}{2} - i(k+1))\,,&s\ {\rm even}\,,\\
\prod_{i=1}^{s/2}(p - \frac{k}{2} + (i-\half)(k+1)) 
(p -\frac{k}{2} - (i-\half)(k+1))\,,&s\ {\rm odd}
\EA\right.
\EA\label{mffr0}\EE
and similarly for $\ket{{\rm MFF}(0,s, p, k)}^-$.
\end{rem}


\subsubsection*{`Exceptional' points}\lvm
A peculiarity of the $\SSL21$ case, not seen in the standard MFF
construction is that, essentially due to the presence of fermions, the
MFF formulae might vanish at certain points in the $p, j,k$ parameter
space. That this is possible can be seen from {\bf\ref{modified}} and
the observations made before Eq.~\req{both}: one might happen to be
unable to return from the modified \hw{} conditions to the standard
ones.

Recall first of all that, as we saw in \req{both}--\req{minusj}, the
MFF singular vector can be alternatively defined with the fermions
$E^1$ and $F^1$ replaced by $E^2$ and $F^2$ respectively, e.g., 
\BE\!\!\new\BA{rcl}
\ket{{\rm MFF}(r,s,p,k)}^+\kern-8pt&=&\kern-8pt
E^2_{-{1\over2}}(F^{12}_0)^{r+(s-1)(k+1)}F^2_{{1\over2}}\cdot
(E^{12}_{-1})^{r+(s-2)(k+1)}\cdot\ldots\cdot
E^2_{-{1\over2}}(F^{12}_0)^{r-(s-1)(k+1)}F^2_{{1\over2}}\cdot{}\\
{}&{}&\hfill{}\cdot\ket{p,\frac{r}{2}-\frac{s-1}{2}(k+1),k}
\EA\label{mffplusother}\EE
As we have remarked, dropping an arbitrary number of factors
from the left of the MFF formulae does still produce a state
which, while not being a Verma module element, does formally
satisfy the \hw{} conditions. 
Let $j_i$, $i\geq1$, be the spin 
(the eigenvalue of $\Hminus_0$) of such a state obtained by keeping
$i-1$ factors acting on the \hw{} vector:
\BE
j_i=\left\{\kern-4pt\new\BA{rl}
\frac{r}{2}-\frac{s-i}{2}(k+1)\,,&i~{\rm odd}\,,\\
-\frac{r}{2} +\frac{s-i+1}{2}(k+1)\,,&i~{\rm even}
\EA\right.
\EE
-- we will thus continue with the MFF$^+$ case, the analysis for
MFF$^-$ is completely similar.

Now, the modified \hw{} conditions \req{hwmod2} are no longer
equivalent to the standard ones as soon as $p-\frac{k}{2}\pm j =0$,
which might indeed be the case with one of the $j_i$. Inside the
`continued' formula, however, this does not necessarily imply the
vanishing of the MFF monomial, since one can then use the other
expression, Eq.~\req{mffplusother}, for the same singular vector.
\label{subtle} 
A different situation occurs when the respective topological \hw{}
conditions are encountered inside \req{mffplusother} as well as in the
formula~\req{mff} for the same vector.  This can happen for $j_{i_1}$
and $j_{i_2}$ with $i_1$ and $i_2$ either simultaneously odd or
simultaneously even (the case when one is odd and the other is even
implies $i_2=i_1+1$ and does not lead to vanishing). Let, for
definiteness, $i_1=2m-1$ and $i_2=2n+1$ where $1\leq m\leq n\leq
s$. This implies
$$
k+1 = {r\over s - m - n}
$$
and therefore the MFF monomial has the following structure (we omit the
\hw{} state for brevity):
\BE
\ldots\,
\underbrace{
E^2_{-{1\over2}}(F^{12}_0)^{\frac{r(n+1-m)}{s - m - n}}F^2_{{1\over2}}
}_{(2n+1){\rm th}}\cdot
(E^{12}_{-1})^{\frac{r(n-m)}{s - m - n}}\cdot\ldots\cdot
(E^{12}_{-1})^{\frac{r(m-n)}{s - m - n}}\cdot
\underbrace{
E^2_{-{1\over2}}(F^{12}_0)^{\frac{r(m-n-1)}{s - m - n}}F^2_{{1\over2}}
}_{(2m-1){\rm th}}\cdot
\ldots\cdot
E^2_{-{1\over2}}(F^{12}_0)^{\frac{r(1-m-n)}{s-m-n}}F^2_{{1\over2}}
\EE
Here, the factors $(2n+1)$th, \ldots, $(2m-1)$th make up the MFF
`singular vector' ${\rm MFF}^+(0, S,
p=\frac{(r+1)m+(1-r)n-s}{2(s-m-n)}, k={r\over s - m - n}-1)$, where
$$
S=n-m+2\,.
$$
Such vectors evaluate as in \req{mffr0}, which does give a vanishing result
for the values of the parameters that we actually have. Analyzing 
similarly the other cases, we arrive at
\begin{thm}\label{vanish}
The MFF singular vectors MFF$^\pm(r,s,p,k)$, Eqs.~\req{mff}, vanish whenever
$(p,k)=\\(\ptop(r,s,m,n), \ktop(r,s,m,n))$, where
\BE\new\BA{rcl}
\ktop(r,s,m,n)+1&=&{r\over s-m-n}\,,\\
\ptop(r,s,m,n)&=&{(r+1)m+(1-r)n-s\over2(s-m-n)}
\EA\qquad\left\{\new\BA{l}
1\leq m\leq s\,,\\
0\leq n\leq s-1\,,\\
m+n\neq s\,,\\
s\geq2\,,\EA.\right. 
\label{exceptional}
\EE
\end{thm}

\begin{rem}
These formulae also guarantee that the charged singular vectors of
both types, I and~II, exist simultaneously with one of the MFF
singular vectors; indeed, with the latter chosen to be ${\rm
MFF}^+(r,s)$, the charged-I and charged-II singular vectors are
labelled by integers $1-m$ and $n$ respectively, as follows from the
fact that Eqs.~\req{exceptional} satisfy the equations
\BE\new\BA{rcl}
p&=&j+\frac{k}{2} -(1-m)(k+1)\,,\\
p&=&-j+\frac{k}{2} -n(k+1)\,,\\
j&=&\frac{r}{2}-\frac{s-1}{2}(k+1)\,.
\EA\EE
(Another solution, $p=\half((s-2n)(k+1)-1)$, $j=\half(1-s)(k+1)$,
$r=0$, we drop in view of~\req{mffr0}. Note however that
\req{exceptional} imposes restrictions on the range of $m$ and $n$,
for which the MFF vectors vanish; this does in no way follow from the
mere fact of a simultaneous existence of several singular vectors.)
\end{rem}
Given \req{exceptional}, we can indeed see that the topological
conditions $p \mp j - \half k=0$, discussed in~\reb{modified}, will
indeed be satisfied by one of the `truncated' MFF states, since
\BE
p-\frac{k}{2}-j_i=\left\{\kern-4pt\new\BA{ll}
\frac{r(2m-i-1)}{2(s-m-n)}\,,&i~{\rm odd}\,,\\
\frac{r(i-2n-2)}{2(s-m-n)}\,,&i~{\rm even}
\EA\right.\qquad
p-\frac{k}{2}+j_i=\left\{\kern-4pt\new\BA{ll}
\frac{r(i-2n-1)}{2(s-m-n)}\,,&i~{\rm odd}\,,\\
\frac{r(2m-i)}{2(s-m-n)}\,,&i~{\rm even}
\EA\right.\qquad
\EE
would necessarily vanish as $i$ runs from 1 to $2s-1$ for the MFF factors.
For a fixed pair $(r,s)$ we thus have a set of $s^2-s$ `exceptional'
points labelled by the integers $m$ and $n$ in the specified range.

It is possible now to define the MFF singular vectors at these 
vanishing points as
\BE\new\BA{rcl}
\ket{{\sf mff}(r,s,m,n,\alpha)}^\pm &=&
\lim_{\epsilon\to0}\left(\frac{1}{\epsilon}
\ket{{\rm MFF}(r,s,\ptop(r,s,m,n)+\epsilon\cos\alpha,
\ktop(r,s,m,n)+\epsilon\sin\alpha)}^\pm\right)\\
{}&=&\left.{\d\over\d\epsilon}\left(
\ket{{\rm MFF}(r,s,\ptop(r,s,m,n)+\epsilon\cos\alpha,
\ktop(r,s,m,n)+\epsilon\sin\alpha)}^\pm\right)\right|_{\epsilon=0}
\EA\label{double}\EE
The result is $\alpha$-dependent, and therefore, for either the
$\ket{~}^+$ or the $\ket{~}^-$ case, there will in fact be {\it
two\/} linearly-independent singular vectors with identical quantum
numbers\,\footnote{The double multiplicity of singular vectors was
first observed in~\cite{[Doerr2]}, and for the $\SSL21$ algebra,
in~\cite{[BT]}.}.

\section{The $\SSL21\longleftrightarrow\N2$ relation}\lvm
In this section we consider how a representation of the affine
$\SSL21$ algebra can be constructed starting with the $\N2$
superconformal algebra, and study some of its properties.

As we have mentioned, the Hamiltonian reduction of $\SSL21$ yields the
$\N2$ superconformal algebra \cite{[BO],[BLNW],[IK]}.  The `inverse'
construction, that of the $\SSL21$ {\it currents\/} in terms of the $\N2$
algebra currents and some free fields, has been given in~\cite{[S-sl21]};
however, the related construction for the \hw{} states was only
outlined in~\cite{[S-sl21]}, and we are going to consider it in more
detail here. In the next section we will then analyze the
correspondence between singular vectors in the $\SSL21$ and $\N2$ \hw{}
modules.

\subsection{The $\N2$ superconformal algebra and its highest-weight 
modules}\lvm In this subsection we review the properties of the $\N2$
algebra that we will need later on. We will closely follow
ref.~\cite{[ST3]}.  Our analysis of $\SSL21$ will be resumed in
Section~\ref{currents}.

The $\N2$ superconformal algebra, taken in the `twisted'
form~\cite{[Ey],[W-top]}, reads
\BE\new\BA{lclclcl}
\left[\cL_m,\cL_n\right]&=&(m-n)\cL_{m+n}\,,&\qquad&[\cH_m,\cH_n]&=
&\frac{\ctop}{3}m\delta_{m+n,0}\,,\\
\left[\cL_m,\cG_n\right]&=&(m-n)\cG_{m+n}\,,&\qquad&[\cH_m,\cG_n]&=&\cG_{m+n}\,,
\\
\left[\cL_m,\cQ_n\right]&=&-n\cQ_{m+n}\,,&\qquad&[\cH_m,\cQ_n]&=&-\cQ_{m+n}\,,\\
\left[\cL_m,\cH_n\right]&=&\multicolumn{5}{l}{-n\cH_{m+n}+\frac{\ctop}{6}(m^2+m)
\delta_{m+n,0}\,,}\\
\left[\cG_m,\cQ_n\right]&=&\multicolumn{5}{l}{2\cL_{m+n}-2n\cH_{m+n}+
\frac{\ctop}{3}(m^2+m)\delta_{m+n,0}\,,}\EA\qquad m,~n\in\oZ\,,
\label{topalgebra}
\EE
where $\ctop$ is the central charge; $\cL$, $\cH$, $\cQ$ and $\cG$ are
called the Virasoro generators, the U(1) current, the BRST current, and the
spin-2 fermionic current respectively. The spectral flow transform on the
$\N2$ algebra is given by
\BE\new\BA{rclcrcl}
\cL_n&\mapsto&\cL_n+\theta\cH_n+\frac{\ctop}{6}(\theta^2+\theta)
\delta_{n,0}\,,&{}&
\cH_n&\mapsto&\cH_n+\frac{\ctop}{3}\theta\delta_{n,0}\,,\\
\cQ_n&\mapsto&\cQ_{n-\theta}\,,&{}&\cG_n&\mapsto&\cG_{n+\theta}
\EA\label{UN2}\EE

A Verma module over the
algebra~\req{topalgebra} is freely generated from a \hw{} vector
$\ket{h,\ell,k}_{N=2}$ by the generators
\BE
\cL_{-m}\,,~m\in\oN\,,\qquad
\cH_{-m}\,,~m\in\oN\,,\qquad
\cQ_{-m}\,,~m\in\oN_0\,,\qquad
\cG_{-m}\,,~m\in\oN\,,
\label{verma}\EE
while $\ket{h,\ell,k}_{N=2}$ satisfies the following set of
equations:
\begin{eqnarray}
\cQ_{\geq1}\,\ket{h,\ell,k}_{N=2}&=&\cG_{\geq0}\,\ket{h,\ell,k}_{N=2}~{}={}~
\cL_{\geq1}\,\ket{h,\ell,k}_{N=2}~{}={}~
\cH_{\geq1}\,\ket{h,\ell,k}_{N=2}~{}={}~0\label{upper}\\
\cH_0\,\ket{h,\ell,k}_{N=2}&=&h\,\ket{h,\ell,k}_{N=2}\,,\qquad
\cL_0\,\ket{h,\ell,k}_{N=2}~{}={}~\ell\,\ket{h,\ell,k}_{N=2}\,.\label{Cartan}
\end{eqnarray}
and $k$ parametrizes the central charge as $\ctop=-3-6k$.  The
parameters $h$ and $\ell$ are called the $U(1)$ charge and dimension
respectively.

Just as it was the case with the $\SSL21$ algebra, the $\N2$ superconformal
algebra does allow for the construction of extremal states,
and these turn out to be important in its representation theory.
For $\ell\neq0$, the state $\ket{h,\ell,k}_{N=2}$ has a `superpartner'
$\ket{h-1,\ell,k;1}_{N=2}=\cQ_0\,\ket{h,\ell,k}_{N=2}$. The state
$\ket{~~;1}_{N=2}$ 
belongs to the same module because
$\ket{h,\ell,k}_{N=2}=1/\ell\,\cG_0\,\ket{h,\ell,k;1}_{N=2}$.  We can
in fact continue acting with the modes of $\cG$ or $\cQ$, each time
with the highest of those modes that do not annihilate the state.  We
thus arrive at the {\it extremal\/} vectors \ldots $E_{-2}$, $E_{-1}$,
$E_{0}$, $E_{1}$, $E_{2}$, $\ldots$, as
\BE
\unitlength=1.00mm
\begin{picture}(140,38)
\put(50.00,04.00){
\put(00.00,00.00){$\bullet$}
\put(10.00,20.00){$\bullet$}
\put(10.00,20.00){$\bullet$}
\put(20.00,30.00){$\bullet$}
\put(30.00,30.00){$\bullet$}
\put(40.00,20.00){$\bullet$}
\put(50.00,00.00){$\bullet$}
\put(01.00,03.00){\vector(1,2){8}}
\put(10.50,18.50){\vector(-1,-2){8}}
\put(11.50,23.00){\vector(1,1){7}}
\put(19.70,29.00){\vector(-1,-1){7}}
\put(22.20,31.50){\vector(1,0){7}}
\put(28.70,29.95){\vector(-1,0){7}}
\put(33.00,30.00){\vector(1,-1){7}}
\put(39.00,22.00){\vector(-1,1){7}}
\put(43.00,19.00){\vector(1,-2){8}}
\put(49.50,03.00){\vector(-1,2){8}}
\put(00.00,13.00){${}_{\cQ_2}$}
\put(07.00,07.00){${}^{\cG_{-2}}$}
\put(09.00,26.50){${}_{\cQ_{1}}$}
\put(16.00,21.00){${}^{\cG_{-1}}$}
\put(23.90,33.50){${}_{\cQ_{0}}$}
\put(24.50,25.50){${}^{\cG_{0}}$}
\put(37.00,27.50){${}_{\cQ_{-1}}$}
\put(32.50,22.00){${}^{\cG_{1}}$}
\put(47.00,13.00){${}_{\cQ_{-2}}$}
\put(41.00,07.00){${}^{\cG_{2}}$}
\put(-08.00,00.00){$E_{2}$}
\put(03.00,21.00){$E_{1}$}
\put(15.00,32.00){$E_{0}$}
\put(32.50,32.00){$E_{-1}$}
\put(42.00,21.00){$E_{-2}$}
\put(52.00,00.00){$E_{-3}$}
\put(00.50,-06.00){$\vdots$}
\put(50.50,-06.00){$\vdots$}
}
\end{picture}
\label{newdiagram}\EE
The commutation relations \req{topalgebra} allow us to travel over the
set of the extremal vectors: up to a scalar factor, every mapping
between two adjacent extremal vectors can be inverted by acting with
the opposite mode of the other fermion, provided the respective scalar
factor does not vanish.  We will in fact relabel these states as {\it
generalized \hw{} states\/} $\ket{h,\ell,k;\theta}_{N=2}$, which
satisfy (with $\ell$ and $h$ measuring the eigenvalues of $\cL_0$ and
$\cH_0$ respectively, although are not identical to them for
$\theta\neq0$, see~\cite{[ST3]})
\BE\new\BA{ll}\new\BA{rclcrcl}
\cL_m\ket{h,\ell,k;\theta}_{N=2}&=&0\,,\quad m\geq1\,,\quad&
\cQ_\lambda\ket{h,\ell,k;\theta}_{N=2}&=&0\,,
&\lambda=-\theta+p\,,\quad p=1,2,\ldots\\
\cH_m\ket{h,\ell,k;\theta}_{N=2}&=&0\,,\quad m\geq1\,,&
\cG_\nu\ket{h,\ell,k;\theta}_{N=2}&=&0\,,&\nu=\theta+p\,,\quad p=0,1,2,\ldots
\EA&\theta\in\oZ\EA\label{integertheta}\EE

The choice of a particular representative with $\theta=0$ in
\req{upper} is merely a convention as long as none of the scalar
factors mentioned above vanishes. When one of them does, the
corresponding extremal state $\ket{h,k;\theta}_{\rm top}$ satisfies
stronger \hw{} conditions,
\BE\new\BA{rclcrcl}
\cL_m\ket{h,k;\theta}_{\rm top}&=&0\,,\quad m\geq1\,,\quad&
\cQ_\lambda\ket{h,k;\theta}_{\rm top}&=&0\,,
&\lambda\in-\theta+\oN_0\\
\cH_m\ket{h,k;\theta}_{\rm top}&=&0\,,\quad m\geq1\,,&
\cG_\nu\ket{h,k;\theta}_{\rm top}&=&0\,,&\nu=\theta+\oN_0
\EA\quad\theta\in\oZ\,;
\label{gentophwint}\EE
which will be called the (generalized) {\it topological\/} \hw{}
conditions (with only two parameters out of $h,\ell,k$ remaining,
since the dimension is already fixed by~\req{gentophwint},
see~\cite{[ST3]} for the details).  The diagram 
represents the case when this happens at the point $E_{-3}$, with
$\cG_2E_{-3}=0$:
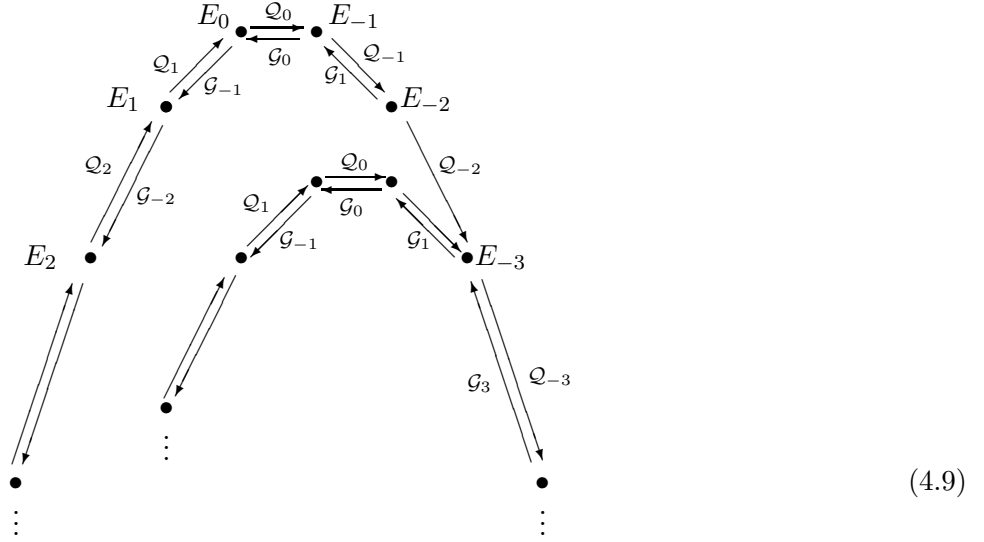
\BE
\unitlength=1.00mm
\begin{picture}(140,66)
\put(30.00, 00.00){
\put(00.00,00.00){$\bullet$}
\put(10.00,30.00){$\bullet$}
\put(20.00,50.00){$\bullet$}
\put(20.00,50.00){$\bullet$}
\put(30.00,60.00){$\bullet$}
\put(40.00,60.00){$\bullet$}
\put(50.00,50.00){$\bullet$}
\put(60.00,30.00){$\bullet$}
\put(70.00,00.00){$\bullet$}
\put(50.00,40.00){$\bullet$}
\put(40.00,40.00){$\bullet$}
\put(30.00,30.00){$\bullet$}
\put(20.00,10.00){$\bullet$}
\put(11.00,33.00){\vector(1,2){8}}
\put(20.50,48.50){\vector(-1,-2){8}}
\put(00.50,03.50){\vector(1,3){8}}
\put(09.90,27.50){\vector(-1,-3){8}}
\put(21.50,53.00){\vector(1,1){7}}
\put(29.70,59.00){\vector(-1,-1){7}}
\put(32.20,61.50){\vector(1,0){7}}
\put(38.70,59.95){\vector(-1,0){7}}
\put(43.00,60.00){\vector(1,-1){7}}
\put(49.00,52.00){\vector(-1,1){7}}
\put(53.00,49.00){\vector(1,-2){8}}
\put(63.00,28.00){\vector(1,-3){8}}
\put(69.50,03.80){\vector(-1,3){8}}
\put(59.30,31.00){\vector(-1,1){7.7}}
\put(52.50,39.40){\vector(1,-1){7.7}}
\put(42.10,41.60){\vector(1,0){8}}
\put(49.55,39.98){\vector(-1,0){8}}
\put(31.80,32.50){\vector(1,1){8}}
\put(40.20,39.00){\vector(-1,-1){8}}
\put(30.20,28.50){\vector(-1,-2){8}}
\put(20.70,12.30){\vector(1,2){8}}
\put(10.00,43.00){${}_{\cQ_2}$}
\put(17.00,37.00){${}^{\cG_{-2}}$}
\put(19.00,56.50){${}_{\cQ_{1}}$}
\put(26.00,51.50){${}^{\cG_{-1}}$}
\put(33.90,63.50){${}_{\cQ_{0}}$}
\put(34.50,55.50){${}^{\cG_{0}}$}
\put(47.00,58.00){${}_{\cQ_{-1}}$}
\put(42.00,53.00){${}^{\cG_{1}}$}
\put(57.00,43.00){${}_{\cQ_{-2}}$}
\put(69.00,15.00){${}_{\cQ_{-3}}$}
\put(61.00,12.00){${}^{\cG_{3}}$}
\put(53.00,31.00){${}^{\cG_{1}}$}
\put(44.00,35.50){${}^{\cG_{0}}$}
\put(44.00,43.50){${}_{\cQ_{0}}$}
\put(31.00,38.00){${}_{\cQ_{1}}$}
\put(36.00,31.00){${}^{\cG_{-1}}$}
\put(02.00,30.00){$E_{2}$}
\put(13.00,51.00){$E_{1}$}
\put(25.00,62.00){$E_{0}$}
\put(42.50,62.00){$E_{-1}$}
\put(52.00,51.00){$E_{-2}$}
\put(62.00,30.00){$E_{-3}$}
\put(00.50,-06.00){$\vdots$}
\put(70.50,-06.00){$\vdots$}
\put(20.50,04.00){$\vdots$}
}
\end{picture}
\label{branchdiag}\EE

\bigskip

\noindent
Then we can further act upon that state with $\cG_{1}$, $\cG_{0}$,
$\cG_{-1}$, \ldots, always preserving the conditions
\req{integertheta}.  We thus see that the extremal diagram
\req{newdiagram} branches at the point where the extra conditions
$\cG_{-r}E_{r-1}=0$ are satisfied.

The crucial point is that the inner parabola in \req{branchdiag}
corresponds to an $\N2$ {\it subrepresentation\/}.  As is elementary
to show, this happens at the $r$th state whenever $\ell=\ell_{\rm
ch}(r,h,k)$, where
\BE
\ell_{\rm ch}(r, h, k)=
r[(r-1)(k+1) + h]\qquad r\in\oZ\,.
\label{ellchN2}\EE
This is the condition for the `charged' series of singular vectors 
to exist~\cite{[BFK]}.  The corresponding singular vector
$\ket{E(r,h,k)}_{\rm ch}$ satisfies the highest-weight conditions
\req{gentophwint} with $\theta=-r$ and is given by
\BE
\ket{E(r,h,k)}_{\rm ch}=
\left\{\new\BA{ll}
\cQ_{r}\,\ldots\,\cQ_0\,\ket{h,\ell_{\rm ch}(r,h,k),k}_{N=2}\,,&r\leq-1\,,\\
\cG_{-r}\,\ldots\,\cG_{-1}\,\ket{h,\ell_{\rm ch}(r,h,k),k}_{N=2}\,,&r\geq1\,,
\EA\right.
\label{thirdplusT}
\label{thirdminusT}
\EE
The minimal-level representative $\ket{S(r,h,k)}_{\rm ch}$ (i.e., the
one at the top of the inner parabola) can be constructed as
\BE
\ket{S(r,h,k)}_{\rm ch}=\left\{
\new\BA{ll}
\cG_0\,\ldots\,\cG_{-r-1}\,\ket{E(r,h,k)}_{\rm ch}\,,
&r\leq-1\,,\\
\cQ_1\,\ldots\,\cQ_{r-1}\,\ket{E(r,h,k)}_{\rm ch}\,,
&r\geq1\,.
\EA\right.
\label{SchN2}\EE
Its level is equal to $|r|$ and the relative charge is $\pm1$ and in
fact equals $r/|r|$ (whence the name, {\it charged\/} singular vectors). It
satisfies the same highest-weight conditions as those we had imposed on the
highest-weight states in~\req{upper}.  

The case $r=0$ is somewhat special because then the topological \hw{}
conditions are satisfied already at the top of the parabola
\req{newdiagram}.  
One thus considers the \hw{} 
vectors $\ket{h,k}_{\rm top}$ that satisfy the topological 
\hw{} conditions \req{gentophwint} for $\theta=0$.
The singular vectors that can be defined in modules 
over $\ket{h,k}_{\rm top}$ 
occur whenever the $U(1)$ charge $h$ of the \hw{} state is one of
the following~\cite{[ST2]}:
\BE\new\BA{rcl}
\htop^+(r,s,k)&=&-(r-1)(k+1)+s-1\,,\\
\htop^-(r,s,k)&=&(r+1)(k+1)-s\,,\EA\qquad r,s\in\oN
\label{hplushminus}\EE
They are given in terms of continued products of the fermionic
generators $g(a,b)\,\mbox{``{}}{=}\mbox{{}''}\,\prod_a^b \cG_\alpha$,
and\\ $q(a,b)\,\mbox{``{}}{=}\mbox{{}''}\,\prod_a^b \cQ_\alpha$
\cite{[ST2],[ST3]}, which we simply quote here in order to be precise
as to the conventions and normalizations when we compare the $\SSL21$
singular vectors with the $\N2$ ones:
\begin{eqnarray}
\ket{E(r,s,k)}^+&=& g(-r,(s-1)t-1)\,
q(-(s-1)t,r-1-t)\,\ldots{}\label{Tplus}\\
{}&{}&\quad{}g((s-2)t-r,t-1)\,
q(-t,r-1-t(s-1))\, g((s-1)t-r,-1)\,\ket{\htop^+(r,s,k),k}_{\rm top}\,,
\nonumber\\
\ket{E(r,s,k)}^-&=&
q(-r, (s-1) t - 1)\,g(-(s-1)t, r - t - 1)\,\ldots\label{Tminus}\\
{}&{}&{}\quad q((s-2) t - r, t-1) \, g(-t, r - (s-1) t - 1)\,
q((s-1) t - r, -1)\,\ket{\htop^-(r,s,k),k}_{\rm top}\nonumber\\
&&r,s\in\oN\nonumber
\end{eqnarray}
where 
$$t\equiv{1\over k+1}\,.
$$
The $\ket{E(r,s,k)}^\pm$ singular vectors are on level $rs+\half
r(r-1)$ over the corresponding topological highest-weight state and
have relative charge $\pm r$.  They satisfy the topological
highest-weight conditions~\req{gentophwint} with~$\theta=\mp r$. The
algebraic rearrangement rules \cite{[ST2]}, which we briefly mentioned
in the Introduction, allow one to rewrite each of these singular
vectors in the `Verma' form, as
$\cE^\pm(r,s,k)\,\ket{\htop^\pm(r,s,k),k}$, where $\cE^\pm(r,s,k)$ are
polynomials in the usual creation operators in the module. To continue
with the remark made after Theorem~\ref{mffthm}, the topological \hw{}
conditions~\req{gentophwint} are fulfilled after the application of
each successive $g$- or $q$- operator; in contrast with a simpler case
of Ka\v c--Moody algebras, however, these conditions hold with a
different $\theta$ at each point.

\medskip

Further, singular vectors may also exist in the modules $U_{h,\ell,k}$
built on $\ket{h,\ell,k}_{N=2}$, $\ell\neq0$; these will be called
`massive' singular vectors. They are given, again, by a continued
construction~\cite{[ST3]} (see also \cite{[Doerr2]} for a different
approach).  In order to fix the relative charge and the level, one can
choose the highest-level representative by fixing $\theta=0$ in the
conditions \req{integertheta}.  Then the singular vectors in
$U_{h,\ell,k}$ are required to satisfy the annihilation conditions
that coincide with those from~\req{upper}.  The relative charge of
these representatives of massive singular vectors vanishes, and the
level is equal to~$rs$.  A massive singular vector exists in
$U_{h,\ell,k}$ if $\ell=\ell(r,s,h)$ for some $r,s\in\oN$,
where~\cite{[BFK]} (for $\ctop\neq3$, i.e. $k\neq-1$)
\BE
\ell(r,s,h,k)
={\left[-(r+1)(k+1)+s+h\right]\left[-(r-1)(k+1)+s-h\right]\over4 (1 + k)}\,,
\quad r,s\in\oN
\label{ellmassiveN2}\EE
Their construction can be given again in terms of `continued products' of
the fermionic generators $\cG$ and $\cQ$ of the algebra, 
$g(a,b)$ and $q(a,b)$ respectively:
\BE\new\BA{l}
\ket{S(r,s,h,k)}_{N=2} = \cN_1(r,s,h,k)\,
g(0, \frac{r-3}{2} +\frac{h+s}{2(k+1)})\\
\quad q(\frac{1-r}{2} -\frac{h+s}{2(k+1)},
\frac{r-1}{2} -\frac{h-s+2}{2(k+1)})\,
g(-\frac{r+1}{2} +\frac{h-s+2}{2(k+1)},
\frac{r-3}{2} +\frac{h+s-2}{2(k+1)})\,\\
\qquad\qquad\ldots\\
\qquad\quad q(\frac{1-r}{2} -\frac{h-s+4}{2(k+1)},
\frac{r-1}{2} -\frac{h+s-2}{2(k+1)})\,
g(-\frac{r+1}{2} +\frac{h+s-2}{2(k+1)},
\frac{r-3}{2} +\frac{h-s+2}{2(k+1)})\,\\
\qquad\qquad q(\frac{1-r}{2} -\frac{h-s+2}{2(k+1)},
\frac{r-1}{2} -\frac{h+s}{2(k+1)})\,
g(-\frac{r+1}{2} +\frac{h+s}{2(k+1)},-1)\,\ket{h,\ell(r,s,h,k),k}_{N=2}
\EA\label{S1}\EE
At the same time, this can be written in a different form,
swapping the roles played by the $q$ and $g$ operators, as
\BE\new\BA{l}
\ket{S(r,s,h,k)}_{N=2} = \cN_2(r,s,h,k)\,
q(1, \frac{r-1}{2} - \frac{h-s}{2(k+1)})\,\\
\quad g(-\frac{r+1}{2} + \frac{h-s}{2(k+1)},
\frac{r-3}{2} + \frac{h+s-2}{2(k+1)})\,
q(\frac{1-r}{2} - \frac{h+s-2}{2(k+1)},
\frac{r-1}{2} - \frac{h-s+2}{2(k+1)})\,\\
\qquad\qquad\qquad\ldots{}\\
\quad\qquad\quad
g(-\frac{r+1}{2} + \frac{h+s-4}{2(k+1)},
\frac{r-3}{2} + \frac{h-s+2}{2(k+1)})\,
q(\frac{1-r}{2} - \frac{h-s+2}{2(k+1)},
\frac{r-1}{2} - \frac{h+s-2}{2(k+1)})\,\\
\quad\qquad\qquad g(-\frac{r+1}{2} + \frac{h+s-2}{2(k+1)},
\frac{r-3}{2} + \frac{h-s}{2(k+1)})\,
q(\frac{1-r}{2} - \frac{h-s}{2(k+1)},0)\,\ket{h,\ell(r,s,h,k),k}_{N=2}
\EA\label{S2}\EE
We have introduced the normalization factors
\BE
\cN_1(r,s,h,k)=\prod_{n=1}^r(h - \eta^+(r, s, 1 - n, k))\,,\qquad
\cN_2(r,s,h,k)=\prod_{n=1}^r(h - \eta^-(r, s, n, k))\,.
\label{Sdata}\EE
with
\BE
\eta^+(r, s, p, k) = s + (-r + 1 - 2 p)(k+1)\,,\qquad
\eta^-(r, s, p, k) = -s + (r + 1 - 2 p)(k+1)
\label{etaplusetaminus}
\EE

The above applies to the generic case, when none of the normalization
factors \req{Sdata} vanishes. 
When {\it both normalization factors \req{Sdata}
vanish\/}, which occurs when
\BE\new\BA{rcl}
\eta^+(r,s,-n,k) = \eta^-(r,s,m,k)\,, \qquad
0 \leq n \leq r-1,\quad 1 \leq m \leq r\,,\quad m+n-r\neq0\,,\quad
r\geq2
\EA\label{special}\EE
that is,
\BE
h = \frac{(1 - m + n)s}{-m - n + r}\,,\qquad 
k+1 = \frac{s}{-m - n + r}\,,
\label{therules}
\EE
none of the formulae \req{S1}, \req{S2} can be applied literally, in
view of the vanishing norm. 
In fact, while the vector
$S(r,s,h,k)\equiv\cN_1(r,s,h,k)\,S^{(1)}(r,s,h,k)=
\cN_2(r,s,h,k)\,S^{(2)}(r,s,h,k)$ vanishes at the points
\req{therules}, the `unnormalized' vectors $S^{(1),(2)}(r,s,\frac{(1 -
m + n)s}{-m - n + r},\frac{s}{-m - n + r}-1)$
provide a basis of a two-dimensional space of singular vectors with
identical quantum numbers at each of the points~\req{therules}. 
This is a preferred basis, since both vectors do then explicitly
factorize through the corresponding {\it
topological\/} singular vector in the centre and the products of the
fermionic modes on the ends of the formulae:
\BE
\kern-9pt\new\BA{rcl}
\ket{\bar {\ssf s}(r,s,m,n)}^{(1)}\kern-6pt&=&\kern-6pt
\Bigl(\prod_{i=0}^{-m + r - 1}\!\!\cG_i\Bigr)\,
\cE^{-,-m}(r,s,\frac{s}{-m - n + r}-1)
\Bigl(\prod_{j=-m}^{-1}\cG_j\Bigr)\,
\ket{\frac{(1 - m + n)s}{-m - n + r},
\frac{m n s}{-m - n + r},\frac{s}{-m - n + r}-1}_{N=2}\,,\\
\ket{\bar {\ssf s}(r,s,m,n)}^{(2)}\kern-6pt&=&\kern-6pt
\Bigl(\prod_{i=1}^{-n + r - 1}\!\!\cQ_i\Bigr)\,
\cE^{+,n}(r,s,\frac{s}{-m - n + r}-1)\,
\Bigl(\prod_{j=-n}^{0}\cQ_j\Bigr)\,
\ket{\frac{(1 - m + n)s}{-m - n + r},\frac{m n s}{-m - n + r},
\frac{s}{-m - n + r}-1}_{N=2}
\EA\label{Sfactor}\EE
where $\cE^\pm(r, s, k)$ are the operators corresponding to the
topological singular vectors \req{Tplus}, \req{Tminus}, and
$\cE^{\pm,\theta}$ denotes the spectral flow transform of $\cE^{\pm}$.
Thus, as soon as the expressions for the topological singular vectors
are known~\cite{[ST2]}, the states \req{Sfactor} 
follow by the above, quite straightforward, construction.
The positive integers $m$ and $n$ measure the distance along the
diagram \req{newdiagram}, from $E_0$ to two branching points, one on
the left and the other on the right half of the parabola
\req{newdiagram}. As a simple example consider the case $r=2$, $s=1$,
$m=2$, $n=1$ (vectors at level $rs=2$)~\cite{[ST3]}:
\BE
\unitlength=1.00mm
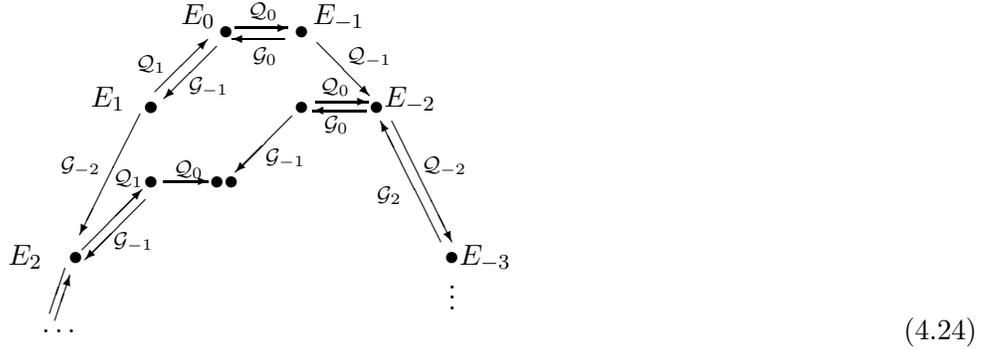
\begin{picture}(160,46)
\put(50.00,10.00){
\put(00.00,00.00){$\bullet$}
\put(10.00,20.00){$\bullet$}
\put(10.00,20.00){$\bullet$}
\put(20.00,30.00){$\bullet$}
\put(30.00,30.00){$\bullet$}
\put(40.00,20.00){$\bullet$}
\put(50.00,00.00){$\bullet$}
\put(09.50,20.00){\vector(-1,-2){8}}
\put(11.50,23.00){\vector(1,1){7}}
\put(19.70,29.00){\vector(-1,-1){7}}
\put(22.20,31.50){\vector(1,0){7}}
\put(28.70,29.95){\vector(-1,0){7}}
\put(33.00,29.50){\vector(1,-1){7}}
\put(43.00,19.00){\vector(1,-2){8}}
\put(49.50,03.00){\vector(-1,2){8}}
\put(-1.00,13.00){${}_{\cG_{-2}}$}
\put(09.00,26.50){${}_{\cQ_{1}}$}
\put(16.00,21.50){${}^{\cG_{-1}}$}
\put(23.90,33.50){${}_{\cQ_{0}}$}
\put(24.50,25.50){${}^{\cG_{0}}$}
\put(37.00,27.50){${}_{\cQ_{-1}}$}
\put(47.00,13.00){${}_{\cQ_{-2}}$}
\put(41.00,07.00){${}^{\cG_{2}}$}
\put(06.00,09.30){${}^{\cQ_1}$}
\put(06.00,01.00){${}^{\cG_{-1}}$}
\put(01.80,02.00){\vector(1,1){8}}
\put(10.20,08.60){\vector(-1,-1){8}}
\put(10.00,10.00){$\bullet$}
\put(12.60,11.00){\vector(1,0){6}}
\put(18.80,10.00){$\bullet\bullet$}
\put(26.20,12.00){${}^{\cG_{-1}}$}
\put(39.50,20.40){\vector(-1,0){7}}
\put(32.80,21.60){\vector(1,0){7}}
\put(30.00,20.00){$\bullet$}
\put(29.70,19.70){\vector(-1,-1){7.5}}
\put(33.50,23.40){${}_{\cQ_0}$}
\put(34.00,18.40){${}_{\cG_0}$}
\put(14.00,12.50){${}_{\cQ_0}$}
\put(-08.00,00.00){$E_{2}$}
\put(03.00,21.00){$E_{1}$}
\put(15.00,32.00){$E_{0}$}
\put(32.50,32.00){$E_{-1}$}
\put(42.00,21.00){$E_{-2}$}
\put(52.00,00.00){$E_{-3}$}
\put(50.50,-06.00){$\vdots$}
\put(-0.50,-0.50){\line(-1,-3){2}}
\put(-1.80,-7.40){\vector(1,3){2}}
\put(-3.50,-09.00){$\ldots$}
}
\end{picture}
\kern-260pt\new\BA[t]{rcl}
\mbox{}\\
\ket{\bar {\ssf s}(2,1,2,1)}^{(1)}&=&\cQ_0\,\cQ_1\,\cG_{-2}\,\cG_{-1}\,
\ket{0,-2,-2}_{N=2}\,,\\
\ket{\bar {\ssf s}(2,1,2,1)}^{(2)}&=&\cG_{-1}\,\cG_0\,Q_{-1}\,\cQ_0\,
\ket{0,-2,-2}_{N=2}
\EA\label{degendiagram}\EE
This completes our review of the $\N2$ algebra and its singular vectors.

\subsection{Constructing the $\SSL21$ currents\label{currents}}\lvm
The affine $\SSL21$ algebra can be constructed in terms of an arbitrary
$\N2$ superconformal matter (where, given the commutation relations
\req{topalgebra}, we introduce the currents
$\cT(z)=\sum_{n\in\oN}\cL_n\,z^{-n-2}$,
$\cG(z)=\sum_{n\in\oN}\cG_n\,z^{-n-2}$,
$\cQ(z)=\sum_{n\in\oN}\cQ_n\,z^{-n-1}$, and
$\cH(z)=\sum_{n\in\oN}\cH_n\,z^{-n-1}$), 
two free bosonic currents with opposite
signatures and a free fermion $\psi\bar\psi$ with the operator
products
\BE
\d F(z)\d F(w)={-k/2\over(z-w)^2}\,,\qquad
\d U(z)\d U(w)={k/2\over(z-w)^2}\,,\qquad
\psi(z)\,\bar\psi(w)={1\over z-w}\,.
\label{DFDU}\EE
Namely,
\begin{thm}[\cite{[S-sl21]}]\label{sl21thm}
Let the $\N2$ central charge be $\ctop = -3 -6 k$, $k\neq0$, and the
free fields $\d F$, $\d U$ and $\psi,\bar\psi$ satisfy the operator
products~\req{DFDU}. Then the currents
\BE\new\BA{rcl}
E^1 &=& \psi\, e^{{1\over k}(U-F)}\,,\qquad\qquad
E^2\;{}={}\,\barpsi\, e^{{1\over k}(U-F)}\,,\qquad
E^{12}\,{}={}\,e^{{2\over k}(U-F)}\,,\\
\Hplus&=&-\half \cH + \half \psi\,\barpsi\,,\qquad
\Hminus = \d U\,,\\
F^1 &=& \bigl(\cG - \barpsi\,\d F  -
\half\cH\,\barpsi -
(k+\half)\d\barpsi\bigr)e^{-{1\over k}(U-F)}\,,\\
F^2 &=& \bigl(\half(k+1)\cQ + \psi\,\d F -
\half\cH\,\psi  +
(k+\half)\d\psi\bigr)e^{-{1\over k}(U-F)}\,,\\
F^{12} &=&
\bigl(-\d F\,\d F - (k+1) \d\d F + (k+2)T_{N=0}\bigr)e^{-{2\over k}(U-F)}\,,
\EA\label{eco}\EE
where
\BE\new\BA{rcl}
T^{{\phantom{y}}}_{N=0}{}&=&{}
\frac{1}{k+2}\Bigl((k+1)(\cT + \half\d\cH) + \fourth\cH\,\cH + \cG\,\psi -
  \half(k+1)\cQ\,\barpsi + \half \cH\,\psi\,\barpsi\\{}&{}&{}+
  \fourth(1 + 2k)\psi\,\d\barpsi-
  \fourth(1 + 2 k)\d\psi\,\barpsi)\Bigr)\,,
\EA\label{TN0}\EE
close to the affine $\SSL21$ algebra of level~$k$\,\footnote{In
\req{eco}, the nested normal orderings are assumed from right to left,
as \ ${:}A\;{:}BC{:}\;{:}$. However, in order to make the formulae
shorter, we write the vertex operators $\exp(a(U-F))$ as common
factors.  Thus Eqs.~\req{eco} should be understood by multiplying
every term in the parentheses with the vertex operator and introducing
the normal ordering as explained.}.  {\rm For the Sugawara \emt{}
\req{N2Sug} we find then
\BE
T_{\rm Sug}=\cT+\half\d\cH-\frac{1}{k}\d F\,\d F - \d\d F
+\frac{1}{k}\d U\d U + \half\d\bar\psi\,\psi-\half\bar\psi\,\d\psi
\label{suguntw}\EE
}
\end{thm}
Here, $\cT+\half\d\cH$ is the `{\it un\/}twisted' \emt{} with central
charge $\ctop$, therefore the appearance of $\d\cH$ in \req{suguntw}
is due to our choice of centreless
$\cT$, rather than $\cT+\half\d\cH$, as the
basic field (see the commutation relations~\req{topalgebra}).

\begin{rem}
If the $\d F$ and $\d U$ currents were normalized to $\pm1$ over the
poles in \req{DFDU}, the exponents in \req{eco} would acquire the
factors $\pm\sqrt{{1\over2k}}$, $\pm\sqrt{{2\over k}}$, and the \emt{}
in~\req{suguntw} would take the canonical form. An essential point is
that, in whatever normalization, the currents $\d U$ and $\d F$ have
opposite signatures; the current $\d U - \d F$ is {\it null}.
\end{rem}

\begin{rem}
The construction $T^{{\phantom{y}}}_{N=0}$ can be singled out
in~\req{eco} only for $k\neq-2$; the usefulness of this object is that
it is an \emt, i.e., satisfies the Virasoro algebra~\cite{[S-sl21]}
(with the `bosonic-matter' central charge $13-\frac{6}{k+2}-6(k+2)$). It is
clear, however, that upon substituting $T^{{\phantom{y}}}_{N=0}$ into
the expression for $F^{12}$, the latter is well-defined for $k=-2$ as
well (in fact, $F^{12}$ is {\it determined \/} by $F^1$ and~$F^2$).
\end{rem}

\begin{rem}\label{string}
In the $\N2$ non-critical string theory, one actually
finds~\cite{[S-sl21]} a realization of $\SSL21$ in terms of a slightly
different field content, namely, in addition to the superconformal matter,
a complex Liouville scalar $\d\bar\phi\,\d\phi$ with a superpartner,
$\bar\psi\,\psi$, and the multiplet of fermionic and bosonic ghosts
$b\,c$, $\eta\,\xi$ and $\beta\,\gamma$, $\tbeta\,\tgamma$. In terms
of these\,\footnote{To be precise, the $\N2$ string conventions are
$$
b(z)\,c(w) = \eta(z)\,\xi(w) = {1\over z-w}\,,
\quad
\tbeta(z)\,\tgamma(w) = 
\beta(z)\,\gamma(w) = {-1\over z-w}\,,
\quad
\d\phi(z)\,\d\bar\phi(w) = {-1\over(z-w)^2}\,.
$$}, the $\DF$ and $\DU$ scalars that we had above are expressed
as
\BE\new\BA{rcl}
\DF &=& -\half \Dbarphi - \half (3 + k) \Dphi +
 b\,c + \half \beta\,\gamma + \half \tbeta\,\tgamma + \eta\,\xi\,,\\
\DU &=& -\half \Dbarphi - \half (3 - k) \Dphi +
  b\,c + \half \beta\,\gamma + \half \tbeta\,\tgamma + \eta\,\xi\,.
\EA\label{DUDF}\EE
Substituting this into \req{eco} results in a realization which is
also valid for $k=0$; however, forgetting about the ghosts and
working with $\DF$ and $\DU$ as
independent fields makes the analysis much more compact, and we will
thus proceed with the representation~\req{eco}.
\end{rem}

In the above realization, some properties of the $\SSL21$ algebra are
realized rather naturally:
\begin{lemma}
The spectral flow transform on the $\SSL21$ generators
constructed as in Theorem~\ref{sl21thm} is realized as
\BE\new\BA{rclcrcl}
\cL_n&\mapsto&\cL_n+\theta\cH_n-(k+\half)(\theta^2+\theta)
\delta_{n,0}\,,&{}&
\cH_n&\mapsto&\cH_n-(2k + 1)\theta\delta_{n,0}\,,\\
\cQ_n&\mapsto&\cQ_{n-\theta}\,,&{}&\cG_n&\mapsto&\cG_{n+\theta}\,\\
\psi_n&\mapsto&\multicolumn{5}{l}{\psi_{n-\theta}\,,\qquad
\bar\psi_n\mapsto\bar\psi_{n+\theta}\,,}
\EA\label{U}\EE
where the first two lines represent the $\N2$ spectral flow transform.
\end{lemma}
Thus the transformation is `localized' in the $\N2$ and
$\bar\psi\,\psi$- sectors (to recall the actual string field content,
$\psi$ is the Liouville superpartner); a similar statement is true as
regards the automorphism \req{auto}:
\begin{lemma}
For $k\neq-1$, the automorphism \req{auto} of the $\SSL21$ algebra is
realized for the construction of Theorem~\ref{sl21thm} as
\BE\new\BA{rclrcl}
\cG&\mapsto&-\frac{k+1}{2}\cQ\,,& \cQ&\mapsto&-\frac{2}{k+1}\,\cG\,,\\
\cH&\mapsto&-\cH\,,\\
\psi&\mapsto&\bar\psi\,,&\bar\psi&\mapsto&\psi\,,
\EA\EE
\end{lemma}
where, again, the first two lines are an involutive automorphism of the
$\N2$ algebra.

At the same time, the construction \req{eco} does obviously break the
symmetry under the automorphism~\req{4thorder} by selecting one out of
the two `triangular' subalgebras.

\smallskip

The following technical observation will be quite useful in what follows:
\begin{lemma}
The normal-ordered products of the field operators~\req{eco} evaluate as
\BE\new\BA{rcl}
E^1\,F^2&=&
\half(1 + k) \psi\,\cQ - (1 + k) \d\psi\, \psi\,,\\
E^2\,F^1&=&\bar\psi\,\cG + (1 + k) \d\bar\psi\,\bar\psi
\EA\label{no}\EE
\end{lemma}

\subsection{Constructing $\SSL21$ \hw{} states\label{states}}\lvm
Now, having seen how the $\SSL21$ currents arise on the non-critical
$\N2$ string worldsheet, we can address the problem of how the
$\SSL21$ representation space is prepared by the string. In order to
present the construction of the representation space of $\SSL21$, we
choose a particular vacuum from the family of the $\theta$-vacua
related by the spectral flow transform.  For $\theta=\half$, we will
have the Ramond state
\BE
\ket{p, j, k;\half}\,,
\label{Ramond}\EE
for which the \hw{} conditions \req{hwmassive} become
\BE\new\BA{rclrclrcl}
E^1_{\geq0}\ket{p,j,k;\half}&=&0\,,&
E^2_{\geq0}\ket{p,j,k;\half}&=&0\,,
&E^{12}_{\geq0}\ket{p,j,k;\half}&=&0\,,\\
F^1_{\geq1}\ket{p,j,k;\half}&=&0\,,&
F^2_{\geq1}\ket{p,j,k;\half}&=&0\,,
&F^{12}_{\geq1}\ket{p,j,k;\half}&=&0\,,\\
\multicolumn{9}{c}{
\Hplus_0\ket{p,j,k;\half}~{}={}~(p-\frac{k}{2})\,\ket{p,j,k;\half}\,,
\qquad
\Hminus_0\ket{p,j,k;\half}~{}={}~j\,\ket{p,j,k;\half}\,}
\EA\label{hwmassiveR}\EE
(as before, the conditions \req{hwH}, unaffected by the spectral flow
transform, are understood.)

A Ramond vacuum satisfying the conditions \req{hwmassiveR} can be
constructed by tensoring an appropriate $\N2$ \hw{} state with the
free-field vacua as follows. There are three (besides the level $k$) a priori
parameters: the $U(1)$ charge and the dimension of the $\N2$ \hw{}
state $\ket{h,l,k}$, and the coefficient $a$ in the vertex operator
$e^{a\,(U - F)}$ in the $U$-$F$-sector\,\footnote{One readily sees
that $U$ and $F$ can only appear in the vertex operator in the
combination $U-F$ in order that this vertex operator could be a part
of a primary state with respect to the currents~\req{eco}.}.  The
candidates for the $\SSL21$ \hw{} states thus read
\BE
\ket{k-2p, l, k}_{N=2}\tensor \ket{e^{a\,(U - F)}}
\tensor\ket0_{\psi\barpsi}\,;
\label{candidate}
\EE
yet another (discrete) parameter might seem to come from the possibility of
choosing different fermionic vacua \cite{[FMS]} in the
$\bar\psi\,\psi$ theory, but this is accounted for by the $\SSL21$
spectral flow transform, and thus is not relevant once we are
considering the $\SSL21$ \hw{} states in a fixed (Ramond) sector.
The vacuum $\ket0_{\psi\barpsi}$ in~\req{candidate} is defined by 
\BE
\psi_{\geq1}\,\ket0_{\psi\barpsi}=\bar\psi_{\geq0}\,\ket0_{\psi\barpsi}=0\,,
\EE
for the {\it integer-moded\/} $\psi$ and $\bar\psi$. With some abuse
of notation, $\ket{e^{a\,(U - F)}}$ denotes the primary state that
corresponds to the vertex operator~$e^{a\,(U - F)}$.

\begin{thm}\label{dim}
The state \req{candidate} satisfies the $\SSL21$ \hw{} conditions
\req{hwmassiveR} iff the $\N2$ dimension is $l=\theell(p, j, k)$,
where $j=ka/2$ and
\BE
\theell(p, j, k) = -{(j + p - \frac{k}{2}) (1 - j + \frac{k}{2} + p)\over 
1 + k}\,.
\label{theell}\EE
\end{thm}

Thus the $\SSL21$ \hw{} state in the Ramond sector reads
\BE
\ket{p, j, k; \half} =
\ket{k-2p, \theell(p, j, k), k}_{N=2}\tensor
\ket{e^{2j/k\,(U - F)}}\tensor\ket0_{\psi\barpsi}
\label{themassive}
\EE
It has dimension $\frac{j^2 - (p-{k\over2})^2}{1 + k}$ with
respect to the Sugawara \emt~\req{suguntw},
\BE\new\BA{l}
L^{\rm Sug}_0\, \ket{k-2p, \theell(p, j, k), k}_{N=2}\tensor \ket{e^{2j/k\,(U
- F)}}\tensor\ket0_{\psi\barpsi}\\
\qquad\qquad{}=
\frac{j^2 - (p-{k\over2})^2}{1 + k}\,\ket{k-2p,
\theell(p, j, k), k}_{N=2}\tensor \ket{e^{2j/k\,(U -
F)}}\tensor\ket0_{\psi\barpsi}\,,
\EA\EE
and therefore becomes `massless' precisely when one of the topological
conditions $j=\pm (p-{k\over2})$ holds (see Eqs.~\req{hwtop2}
and~\req{hwtop1}, applied to the state~\req{Ramond}).

\begin{rem}
The spectral flow transform \req{spectral} with arbitrary $\theta$ can
be applied to both sides of \req{themassive}, where on the right-hand
side the $\N2$ spectral flow transform \req{U} will be accompanied by
the `spectral flow transform' on the free fermions (in particular, by
replacing the vacuum $\ket0_{\psi\barpsi}$ with the corresponding
$q$-vacuum~\cite{[FMS]} with $q=\theta$). This gives all the
generalized $\SSL21$ \hw{} states~$\ket{\tilde p, j, k; \theta}$.
\end{rem}

For a given dimension $l$ of the $\N2$ \hw{} state $\ket{h,l,
k}_{N=2}$, and with the $U(1)$ charge fixed as $h=k-2p$, we have
therefore two solutions for the $\SSL21$ spin~$j$:
\BE
j=\half\left(k+1\pm\sqrt{1+4p+4p^2+4(k+1)l}\right)\,.
\EE
Accordingly, there are in general two ways to dress an $\N2$ \hw{} state into
an $\SSL21$ \hw{} state.

\medskip

Now, let us assume that there exists a `charged' singular vector
\req{thirdplusT} over the $\N2$ state $\ket{h,
l, k}_{N=2}$. This means that the dimension $l$ of the $\N2$ \hw{}
state must be of the form $l = \ell_{\rm ch}(r, h, k)$, with $\ell_{\rm ch}$
given by~\req{ellchN2}; setting also $h=k-2p$, we will thus have the
$\N2$ state 
$$\ket{k-2p, r [(r-1)(k+1) -2p + k], k}_{N=2}.$$
Then, let us dress this state into an $\SSL21$ \hw{} state according to
the recipe of Theorem~\ref{dim}. We thus arrive at
\begin{lemma}
Every $\N2$ \hw{} state $\ket{k-2p, \ell_{\rm ch}(r, k-2p, k), k}$,
$r\in\oZ$, can be dressed into an $\SSL21$ \hw{} state
$\ket{p,j,k;\half}$ in the Ramond sector in precisely two ways,
namely into states of the form \req{themassive} with either of the two
values of~$j$:
\BE
j =  p - \frac{k}{2} + (1 - r) (k + 1)\,,\quad {\rm or}\quad
j = -p + \frac{k}{2} + r (k + 1)\,;
\label{jch}
\EE
expressed in terms of $j$ considered as an independent parameter, these
relations reproduce~Eqs.~\req{ptop}:
\BE
p=\ptop_1(1-r, j, k)=j + \frac{k}{2} + (r-1)(k+1)\,,\quad{\rm or}\quad
p=\ptop_2(-r, j, k)=-j + \frac{k}{2} + r(k+1)
\label{pch}\EE
respectively.  Each of the resulting $\SSL21$ \hw{} states is therefore
such that a charged singular vector exists on it.  \rm
\end{lemma}

We can thus expect that the charged $\SSL21$ singular vectors
\req{Ech2} and \req{Ech1} would be related to the charged $\N2$
singular vectors~\req{thirdplusT}. As is clear from the above, the
existence of spectral flow transforms for each of the algebras
involved allows us to consider this only for, say, charged-II $\SSL21$
singular vectors (recall that the charged-I vectors follow by a
combination of the spectral flow transform and the automorphism, {\it
both of which allow for a restriction to the $\N2$ superconformal
algebra\/}).

\medskip

Further, let us take the $\N2$ \hw{} state on which a massive $\N2$
singular vector exists; this determines the $\cL_0$-dimension as
in~\req{ellmassiveN2}, and thus the $\N2$ state under consideration
is~$\ket{k-2p, \ell(r,s,k-2p,k), k}_{N=2}$.  
\begin{lemma}\label{everymassive}
Every $\N2$ \hw{} state $\ket{k-2p, \ell(r, s, k-2p, k), k}$, $r,s\in\oN$, can
be dressed into an $\SSL21$ \hw{} state in the Ramond sector in precisely two
ways, namely into states of the form \req{themassive} with either of the two
values of $j$:
\BE
j = \jminus(s,r+1,k)=- \half s + \half(r + 1)(k+1) \,, \qquad
j = \jplus(s,r,k)=\half s - \half(r - 1)(k+1)
\label{jmff}
\EE 
(cf.~\req{jpm}).  Each of the resulting $\SSL21$ \hw{} states is
therefore such that an MFF singular vector exists on it.
\end{lemma}

As we had $r,s=1,2,\ldots$, in \req{ellmassiveN2}, we thus reproduce
in \req{jmff} all the cases from \req{jpm} except $\jminus(n,1,k)$,
$n\geq1$. These $\SSL21$ \hw{} states do {\it not\/} therefore allow
for a construction in terms of $\N2$ \hw{} states on which an $\N2$
singular vector can 
live\,\footnote{
This fact suggests that the $\SSL21$ algebra would have an additional
series of fusions as compared to the $\N2$ algebra (similarly to an
`extra' series of fusions that exist in the $\SL2$ theory as compared
to the minimal models~\cite{[AY],[Andreev],[Petersen]}).}.

\section{$\SSL21$ singular vectors on the $\N2$ string worldsheet}\lvm
As we have seen in the previous section, every $\N2$ Verma module
$U_{h,\ell,k}$ can be dressed, by tensoring it with free-field modules,
into two $\SSL21$ modules. Further, whenever $U_{h,\ell,k}$ has a singular
vector, the resulting $\SSL21$ module would also have a singular vector.
This applies in fact to the entire embedding diagrams of $\N2$
singular vectors, since the $\N2$ singular vectors, viewed as \hw{} states,
can again be dressed according to the recipe of Theorem~\ref{dim}:
\begin{lemma}
Every $\N2$ singular vector can be dressed as specified in
Theorem~\ref{dim}, into a state that satisfies the $\SSL21$ \hw{}
conditions.
\end{lemma}
Note that this lemma does not tell us anything about whether a given
$\SSL21$ singular vector can be arrived at by dressing an $\N2$
singular vector, nor in fact whether the $\SSL21$ \hw{} state
resulting from the dressing would be an $\SSL21$-descendant of the
chosen \hw{} state in the module.

To see that the above statement is true, we consider first
the charged singular vectors. 
As regards their \hw{} properties, the states \req{SchN2}
can be written as
$\ket{h+r/|r|,\ell_{\rm ch}(r,h,k)+|r|/r,k}_{N=2}$, where 
$\ell_{\rm ch}(r,h,k)+|r|/r=r\left[(r-1)(k+1) + h + r/|r|\right]$,
hence Theorem~\ref{dim} applies with $k-2p\equiv h \leadsto h + r/|r|$. 
As to the massive $\N2$ singular vectors, similarly, we can write them as
$\ket{h,\ell(r,s,h,k)+rs,k}_{N=2}$, where the dimension factorizes as
$$
\ell(r,s,h,k)+rs={[s+r(k+1)-h+(k+1)][s+r(k+1)+h-(k+1)]\over4(k+1)}\,,
$$
therefore Theorem~\ref{dim} would apply again and result in a shift of
$j$.  In fact, in both cases the effective shifts of the resulting
$\SSL21$ parameters are such as they would be if the resulting $\SSL21$
state was the respective (charged, or MFF-) $\SSL21$ singular
vector. As we are going to see, this is indeed the case!

\subsection{The charged singular vectors}\lvm
Now, we proceed to a more complicated problem of a direct evaluation
of $\SSL21$ singular vectors in the $\N2$ terms.  Consider first the
charged singular vectors. We will take them in the `Ramond' sector: 
denote
\BE
\ket{E(r, j, k)}^{(i),\,{\rm (R)}}_{\rm ch}=
\cU_{{1\over2}}\,\ket{E(r, j, k)}^{(i)}_{\rm ch}\,,
\EE
with $\cU_\theta$ being the spectral flow transform operator~\req{spectral}.

Clearly, it suffices to evaluate in the realization \req{eco},\req{themassive} 
the {\it topological representatives\/} of the
charged $\SSL21$ singular vectors~\req{Ech2}, \req{Ech1}. These
become, again, the topological representatives of the charged $\N2$
singular vectors~\req{thirdplusT}:

\begin{thm}\label{chargedred}
The charged-II singular vectors \req{Ech2}, mapped into the Ramond sector,
evaluate in the realization \req{eco}, \req{themassive} as
the charged $\N2$ singular vectors \req{thirdplusT} tensored with
the free-field vacua:
\BE
\ket{E(r, j, k)}^{(2),\,{\rm (R)}}_{\rm ch}=\left\{\!\!\new\BA{l}
\cG_{r}\,\ldots\,\cG_{-1}
\ket{k-2\ptop_2(r,j,k), 
\ell_{\rm ch}(-r,k-2\ptop_2(r,j,k), k), k}_{N=2}\\
\qquad{}\tensor\ket{e^{2j/k\,(U - F)}}\tensor
\bar\psi_{r}\,\ldots\,\bar\psi_{-1}\,
\ket0_{\psi\barpsi}\,,\hfill{}r\leq-1\,,\\
(-1)^r\,\left(\frac{k+1}{2}\right)^{r+1}\,\cQ_{-r}\,\ldots\,\cQ_0
\ket{k-2\ptop_2(r,j,k), 
\ell_{\rm ch}(-r,k-2\ptop_2(r,j,k), k), k}_{N=2}\\
\qquad{}\tensor\ket{e^{2j/k\,(U - F)}}\tensor
\psi_{-r+1}\,\ldots\,\psi_{0}\,
\ket0_{\psi\barpsi}\,,\hfill{}r\geq0
\EA\right.
\EE
{\rm (Of course, $\ell_{\rm ch}(-r,k-2\ptop_2(r,j,k),k)=
\theell(\ptop_2(r,j,k),j,k)$). Here,
$\bar\psi_{r}\,\ldots\,\bar\psi_{-1}\, \ket0_{\psi\barpsi}$ and
$\psi_{-r+1}\,\ldots\,\psi_{0}\, \ket0_{\psi\barpsi}$ represent the `$r$'
vacua in the $\psi\bar\psi$ theory~\cite{[FMS]}.}
\end{thm}

Either applying the spectral flow transform, or directly evaluating the 
charged-I $\SSL21$ singular vectors \req{Ech1}, we arrive at

\renewcommand\thelemma{\thesection.\arabic{lemma}\mbox{$'$}}
\addtocounter{lemma}{-1}
\begin{thm}
The charged-I singular vectors \req{Ech2} mapped into the Ramond sector
evaluate in the realization \req{eco}, \req{themassive} as
the following tensor products of the $\N2$ charged
singular vectors with the free-field vacua::
\BE
\ket{E(r, j, k)}^{(1),\,{\rm (R)}}_{\rm ch}=\left\{\!\!\new\BA{ll}
(-1)^r\,\cG_{r-1}\,\ldots\,\cG_{-1}
\ket{-2\ptop_1(r,j,k), 
\ell_{\rm ch}(1-r,k-2\ptop_1(r,j,k), k), k}_{N=2}\\
\qquad{}\tensor\ket{e^{2j/k\,(U - F)}}\tensor
\bar\psi_{r}\,\ldots\,\bar\psi_{-1}\,
\ket0_{\psi\barpsi}\,,&r\leq0\,,\\
\left(\frac{k+1}{2}\right)^{r}\,\cQ_{-r+1}\,\ldots\,\cQ_0
\ket{-2\ptop_1(r,j,k), 
\ell_{\rm ch}(1-r,-2\ptop_1(r,j,k), k), k}_{N=2}\\
\qquad{}\tensor\ket{e^{2j/k\,(U - F)}}\tensor
\psi_{-r+1}\,\ldots\,\psi_{0}\,
\ket0_{\psi\barpsi}\,,&r\geq1
\EA\right.
\label{chargedIIred}\EE
\end{thm}
\renewcommand\thelemma{\thesection.\arabic{lemma}}

In each case we thus get a charged $\N2$ singular vector tensored with
a $UF$-vertex operator and an $\ket{r}$-vacuum in the
$\bar\psi\,\psi$-theory. 

\begin{rem}
Taking the fields to be literally those of
the non-critical $\N2$ string in the conformal gauge
(see~\req{DUDF}), we would have $e^{2j/k\,(U-F)}=e^{2j\,\phi}$, and
we thus see that, \ i)~the ghosts decouple altogether (the RHSs of
\req{chargedIIred} and similar formulae would then contain only the
product of bare ghost vacua), ii)~the super-Liouville sector, on the
other hand, is sensitive to which singular vector is being evaluated:
it contributes $\ket{e^{2j\,\phi}}\tensor\ket{r}_{\psi\bar\psi}$
to the $\ket{E(r, j, k)}_{{\rm ch}}$ singular vector.
\end{rem}

Being interested in the $\N2$ piece, and thus dropping down the $UF$
and $\psi\bar\psi$ sectors, we can summarize the situation as
the following {\it reductions\/} of singular vectors:
\BE\new\BA{rcccl}
E(r,j,k)^{(2)}_{\rm ch}\kern-18pt&{}&{}&{}&\kern-18pt 
E(r+1,j,k)^{(1)}_{\rm ch}\\
{}&\mbox{\Large$\searrow$}&{}&\mbox{\Large$\swarrow$}&{}\\
{}&{}&\kern-12pt E(-r,k-2\ptop_2(-r,j,k),k)_{\rm ch}\kern-12pt
\EA\EE

One may wish to choose different representatives for the singular
vectors involved in these reductions.  Such a choice would not, of
course, change the fact that the charged singular vectors of the two
algebras are in the $2:1$ correspondence; a simple analysis shows that
the singular vectors \req{ground2} and \req{ground1} correspond
precisely to the $\N2$ singular vectors~\req{SchN2}.

It should be observed (in fact, this underlies the proof of the
Theorem) that, {\sl when evaluating the extremal vectors\/}, the
fermionic $\SSL21$ currents \req{eco} behave in accordance with the
effective replacements
\BE\new\BA{rclcl}
E^1 &=& \psi\, e^{{1\over k}(U-F)}&\to&\psi\crossbox e^{{1\over k}(U-F)}\,,\\
F^2 &=& \bigl(\half(k+1)\cQ + \psi\,\d F -
\half\cH\,\psi  +
(k+\half)\d\psi\bigr)e^{-{1\over k}(U-F)}&\to&\half(k+1)\cQ
\crossbox e^{-{1\over k}(U-F)}\,,\\
E^2 &=& \barpsi\, e^{{1\over k}(U-F)}&\to&\barpsi
\crossbox e^{{1\over k}(U-F)}\,,\\
F^1 &=& \bigl(\cG - \barpsi\,\d F  -
\half\cH\,\barpsi -
(k+\half)\d\barpsi\bigr)e^{-{1\over k}(U-F)}&\to&\cG
\crossbox e^{-{1\over k}(U-F)}\,,
\EA\label{simred}\EE
where $\crossbox$ stresses the fact that the two factors are decoupled,
whence $E^1F^2\to\half(k+1)\psi\,\cQ$, $E^2F^1\to\bar\psi\cQ$.
In the extremal vectors, these generators are indeed encountered in
combinations $E^2_{-n}\,F^1_{-n+1}$ or $E^1_{-n}\,F^2_{-n+1}$, and it
is to these combinations that the above replacement effectively
applies. This is not surprising in view of \req{no}, where the
right-hand sides are in an obvious correspondence with the pairwise
products of the right-hand sides from \req{simred}, modulo the terms
that {\it vanish inside the extremal vectors\/}. Indeed, the extremal
vectors in the $\psi\bar\psi$-sector, for example, rewrite in terms of
the normal products as
$\d^N\psi\,\ldots\,\d\psi\,\psi(z)$,
and therefore, e.g.,
$\d\psi\,(\psi\,\cQ+\alpha\d\psi\,\psi)=\d\psi\,\psi\,\cQ$.

\subsection{The MFF vs. `massive' singular vectors}\lvm
Now we turn to the MFF singular vectors~\req{mff}.  Recall
Lemma~\ref{everymassive}; we are going to formulate the `inverse'
statement.  As before, ${}^{{\rm(R)}}$ will refer to the Ramond
sector.
\begin{thm}\label{mffred}
The singular vectors $\ket{{\rm MFF}^-(r,1,p,k)}$, evaluated in the
realization \req{eco}, \req{themassive}, become
\BE
\ket{{\rm MFF}^-(r,1,p,k)}^{{\rm(R)}}=
\ket{k-2p,\frac{1}{k+1}(\frac{r-1}{2}-p)(\frac{r+1}{2}+p),k}_{N=2}\tensor
\ket{e^{{r + k + 1\over k}\,(U-F)}}\tensor\ket0_{\psi\barpsi}
\quad r\geq1\,.
\EE
\end{thm}
In terms of the $\N2$ algebra, therefore, these $\SSL21$ singular
vectors reduce to \hw{} states, {\it not\/} to a singular vector; the
$U$-$F$- and $\psi\bar\psi$-sectors do decouple however.

Now, as to the remaining MFF singular vectors, we do not have a direct proof,
yet on the basis of various consistency checks and explicit evaluations
we formulate the following\nopagebreak
\begin{thm}{\sc(Conjectured)}\label{mffredmore}\\
I.~The other MFF singular vectors \req{mff} evaluate as the massive $\N2$
singular vectors tensored with free-field primary states:
\begin{eqnarray}
\ket{{\rm MFF}^-(r,s+1,p,k)}^{{\rm(R)}}&=&
(-1)^s\,2^{r-s}\,r\,\left(\frac{k+1}{2}\right)^{rs}\,
\ket{S(s,r,k-2p,k)}_{N=2}
\tensor\ket{e^{{r+s(k+1)\over k}\,(U - F)}}\tensor\ket0_{\psi\barpsi}
\nonumber\\
\ket{{\rm MFF}^+(r,s,p,k)}^{{\rm(R)}}&=&
(-1)^{s-1}\,2^{r+1-s}\,r\,\left(\frac{k+1}{2}\right)^{r(s-1)}
\label{conjectured}\\
{}&{}&{}\qquad{}\times
\ket{S(s,r,k-2p,k)}_{N=2}
\tensor\ket{e^{{-r-(s-1)(k+1)\over k}\,(U - F)}}\tensor\ket0_{\psi\barpsi}
\nonumber\\
{}&{}&\hfill{}r,s\geq1\,.\nonumber
\end{eqnarray}
\end{thm}

Again, in terms of the fields on the $\N2$ string worldsheet, it is
the Liouville dressing of an $\N2$ \hw{} state that is sensitive to
which $\SSL21$ singular vector is taken, while the ghosts contribute
only the bare vacua.

Thus, with the exception of ${\rm MFF}^-(r,1,p,k)$, the MFF singular
vectors \req{mff} are a `double-covering' of the massive $\N2$
singular vectors.  This can be summarized as follows:
\BE\new\BA{crcccl}
\BA{r}{\rm MFF}^-(r,1,p,k)\\
                  r\geq1\EA&\BA{r}{\rm MFF}^-(r,s+1,p,k)\\
                         r,s\geq1\EA\kern-20pt
                                       &{}&{}&{}&\kern-20pt
                                                \BA{r}{\rm MFF}^+(r,s,p,k)\\
                                                                 r,s\geq1\EA\\
\mbox{\Large$\downarrow$}&{}&
\mbox{\Large$\searrow$}&{}&\mbox{\Large$\swarrow$}\\
\bullet&{}&{}&\kern-12pt S(s,r,k-2p,k)\kern-12pt&
\EA\label{mffreddiagr}\EE
The Theorem is conjectured on the basis of `numerology' (matching the
quantum numbers), Lemma~\ref{everymassive}, several consistency checks
and explicit evaluation in the following cases:
\BE\new\BA[t]{rccc}
{}_r\kern-4pt\setminus\rlap{\mbox{\kern-4pt${}^{s+1}$}}\kern-2pt&2&3&4\\
1&8/3_{\mbox{\small \bf 1}}&36/9_{\mbox{\small \bf 2}}&
139/22_{\mbox{\small \bf 3}}\\
2&33/9_{\mbox{\small \bf 2}}&442/51_{\mbox{\small \bf 4}}\\
3&107/22_{\mbox{\small \bf 3}}
\EA\qquad
\new\BA[t]{rcccc}
{}_r\kern-4pt\setminus\rlap{\mbox{\kern-4pt${}^{s}$}}\kern-4pt&1&2&3&4\\
1&2/3_{\mbox{\small \bf 0}}&11/9_{\mbox{\small \bf 1}}&
48/22_{\mbox{\small \bf 2}}&171/51_{\mbox{\small \bf 3}}\\
2&2/9_{\mbox{\small \bf 0}}&49/51_{\mbox{\small \bf 2}}&\\
3&2/22_{\mbox{\small \bf 0}}&\\
4&2/51_{\mbox{\small \bf 0}}&
\EA
\EE
for ${\rm MFF}^-$ and ${\rm MFF}^+$ respectively, where the notation
$m/n$ indicates that the corresponding $\SSL21$ singular vector
contains $m$ terms when rewritten in the Verma form, while the
respective $\N2$ singular vector has $n$ terms, and the subscript
indicates the {\it level\/} of the $\SSL21$ singular vector.  The next
check would be the reduction of $\ket{{\rm MFF}(2,3,p,k)}^+$ to
$\ket{S(3,2,k-2p,k)}_{N=2}$, with $588/221_{\mbox{\small \bf 4}}$, and that
of $\ket{{\rm MFF}(3,2,p,k)}^+$ to $\ket{S(2,3,k-2p,k)}_{N=2}$, with
$161/221_{\mbox{\small \bf 3}}$, but these reductions seem to be too 
complicated to be performed explicitly.

\begin{rem}
The $\SSL21$ and $\N2$ singular vectors that are being compared in
\req{conjectured}, are initially given each in its own monomial form,
and yet the respective normalizations differ only by inessential
numerical factors and powers of $(k+1)$, whose origin is clear
from~\req{eco}. Thus no $p$-dependent factors or other $k$-dependent
factors appear when the $\N2$ singular vectors are taken directly
from~\cite{[ST3]}.
\end{rem}

The `transposition' of $r$ and $s$ observed in \req{mffreddiagr}
does also show up in the `exceptional' cases, when there are two
linearly independent singular vectors with identical quantum numbers.
Recall that, for both the algebras involved, these `special' singular
vectors have been defined by essentially `resolving' a zero of the
general formula. Comparing \req{exceptional} and \req{therules} we see
that, indeed, the sets of `exceptional' points are mapped onto each
other by $r\leftrightarrow s$.  Substituting the respective parameters
into \req{conjectured}, we would have $0=0$, since the
parameters~\req{therules} are indeed recovered as
$h\Bigm|_{\req{therules}}=2\ktop(s,r,m,n)-\ptop(s,r,m,n)$,
$k\Bigm|_{\req{therules}}+1=\ktop(s,r,m,n)+1$; as before, we are
interested in what is `behind' these zeroes.  We thus have the second
part of the theorem: \addtocounter{lemma}{-1}
\begin{thm}\mbox{}\\
II.~For either $\ket{{\sf mff}(r,s,m,n,\alpha)}^+$, $r\geq1,s\geq2$,
or $\ket{{\sf mff}(r,s+1,m,n,\alpha)}^-$, $r,s,\geq1$, the
two-dimensional space of $\SSL21$ singular vectors at the points
\req{exceptional}, defined by Eqs.~\req{double}, reduces to the
two-dimensional space spanned by the $\N2$ singular vectors
$\ket{{\ssf s}(s,r,m,n}^{(1)}$ and $\ket{{\ssf s}(s,r,m,n}^{(2)}$
given by~\req{Sfactor}.
\end{thm}

\medskip

The nature of the correspondence between the $\SSL21$ and $\N2$
singular vectors is such that it extends to the cases when one
singular vector can be constructed on another one, and moreover, a
`synchronous' appearance of fermions in the expressions for singular
vectors of the two algebras indicates that these `composite' singular
vectors would also vanish simultaneously.  It should also be stressed
that, as we see, the $\SSL21$ singular vectors evaluated in the
`$\N2$-string' realization {\it do not\/} coincide; it is only upon
projecting out the $U$-$F$ and $\bar\psi\,\psi$-sectors\,\footnote{In
`stringy' terms, the Liouville sector.} that, in terms of the $\N2$
algebra alone, ${\rm MFF}^-(r,s+1,p,k)$ and ${\rm MFF}^+(r,s,p,k)$,
and similarly $\ket{E(r, j, k)}^{(2)}_{\rm ch}$ and $\ket{E(r+1, j,
k)}^{(2)}_{\rm ch}$, become identical.  Also, none of the $\SSL21$
singular vectors vanishes when evaluated in the $\N2$ string
realization of $\SSL21$, which is a crucial difference from {\it
free\/}-field constructions, which usually lead to the vanishing of a
number of singular vectors.

On the other hand, the above results do also apply to the Wakimoto
representation of $\SSL21$ \cite{[BO]}, or more precisely, {\it two\/}
Wakimoto representations~\cite{[BKT]}, associated with two
Weyl-inequivalent simple root systems. As shown in~\cite{[S-sl21]},
the free-field ingredients of each of these Wakimoto representations
can be constructed from the fields we had in \req{eco}, {\it once the
$\N2$ matter is `bosonized' in terms of free fields\/}.  Therefore the
Wakimoto bosonizations can be {\it mapped through\/} the `stringy'
representation,\footnote{Note, in particular, that it follows
immediately from the formulae of~\cite{[S-sl21]} that each of the two
Wakimoto bosonizations allows a natural realization of the spectral
flow transform.} and thus the $\SSL21$ singular vectors in the
Wakimoto representations become simply the respective `bosonizations'
of the $\N2$ superconformal singular vectors.

\section{Conclusions}\lvm
We have seen that representation theories of the affine $\SSL21$ and
$\N2$ superconformal algebras are closely related, and have
constructed explicit mappings that implement this relation. In
particular, we have found how the general constructions for the affine
$\SSL21$ and $\N2$ superconformal singular vectors are mapped into
each other. This has been done using the representation of the affine
$\SSL21$ algebra realized in the non-critical $\N2$ string. As to the
general construction of singular vectors, we have seen that both the
$\SSL21$ and $\N2$ singular vectors are initially given in {\it
monomial\/} forms, in terms of `continued' objects. It would be
extremely interesting to directly map these monomial forms into each
other.  This is in fact what we have done for the charged series of
singular vectors, but the monomials in that case did not require a
continuation.
We believe that in the `massive' case as well, the correspondence
between the $\SSL21$ and $\N2$ singular vectors exists, at the most
fundamental level, for the respective `continued monomials', since
they in fact represent the continuation of the extremal states, which
encode a significant part of the structure of the algebra and its
representations~\cite{[FS]}.

\medskip

At the same time, the $\N2$ singular vectors can be mapped into the
affine $\SL2$ singular vectors; to be more precise, the subset of
topological $\N2$ singular vectors is {\it isomorphic\/} to the
standard $\SL2$ singular vectors, while the massive $\N2$ singular
vectors are mapped into $\SL2$ modules without a unique \hw{}
vector~\cite{[FST]}. It would be interesting to build a direct
relation between the $\SSL21$ and $\SL2$ singular vectors, and in
particular to see how these more general $\SL2$ `Verma' modules (those
with infinitely-many equivalent `almost-\hw' vectors) can be derived
from the $\SSL21$ Verma modules.

Returning to the role of the $\SSL21$ representation theory in string
theory, it may indeed be the case that, modulo several
exceptions,\footnote{i.e., several series of singular vectors, which
do {\it not\/} reduce; such series are expected to exist on the
general ground that `larger' algebras should have extra series in
their fusion rules, and this is precisely what we have seen in the
$\SSL21\to\N2$ reduction of singular vectors, and also what is the
case for the $\N2\to{\rm Virasoro}$ reduction.} the diagram
\req{sl21diagr} for the singular vectors will be a `double-covering'
of the diagram~\req{sl2diagr}:
\BE
\unitlength=1pt
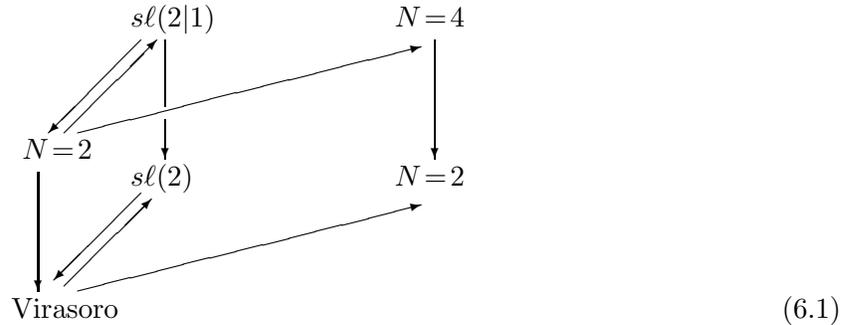
\begin{picture}(250,120)
\put(60,0){Virasoro}
\put(64,60){$\N2$}
\put(70,55){\vector(0,-1){45}}
\put(80,70){\vector(1,1){35}}
\put(109,105){\vector(-1,-1){35}}
\put(105,110){$\SSL21$}
\put(205,110){$\N4$}
\put(85,70){\vector(4,1){130}}
\put(205,50){$\N2$}
\put(85,10){\vector(4,1){130}}
\put(220,105){\vector(0,-1){45}}
\put(80,12){\vector(1,1){33}}
\put(109,47){\vector(-1,-1){33}}
\put(105,50){$\SL2$}
%
\put(118,105){\line(0,-1){25}}
\put(118,75){\vector(0,-1){15}}
\end{picture}
\label{covering}
\EE 
(where an arrow $\cA\to\cB$ means `take generators of $\cA$ and
construct generators of $\cB$, possibly (for the upward arrows) using
some other fields'; of course, the standard notation would be the
embeddings of subalgebras, e.g., $\SL2\hookrightarrow\SSL21$).
The $\N2$ algebra enters this diagram twice, once as an `elementary'
$\N2$ matter theory, and the other time as the algebra realized on the
bosonic string worldsheet.  Accordingly, the affine $\SSL21$ algebra
does also `cover' the non-critical bosonic string~\cite{[BLNW]}. As
regards the proposal to define non-critical string theories as the
Hamiltonian reduction of the appropriate affine
(super)algebras~\cite{[BLNW],[LLS],[RSS]}, one needs to know how much
of the physical content of the worldsheet formulation of a string
theory comes with the Hamiltonian reduction, or at least which
representations of the respective affine superalgebra should be taken
in order to arrive at the string space of states. The present paper
offers a partial result in that direction, by showing how the \hw{}
states are related and also finding out which of the $\SSL21$ singular
vectors do, and which do not, reduce to those of the $\N2$ superconformal
matter theory. There are also some indications that a similar
treatment can be applied to the $\N4$ superconformal algebra, its
\hw{} and extremal states and singular vectors.

\medskip

Now that we have seen that the structure of both the $\SSL21$ and
$\N2$ singular vectors reflects the existence of extremal vectors of
these algebras, an interesting point is how the Lian--Zuckerman states
can also be understood in terms of extremal vectors.  It would also be
very interesting to translate the general constructions and the
reductions of singular vectors into the information about the fusion
rules.

\paragraph{Acknowledgements} It is a pleasure to thank D.~L\"ust
for hospitality at Humboldt-Universit\"at zu Berlin, where this paper
was written.  I am also grateful to O.~Andreev, S.~Ketov,
C.~Preitschopf, and I.~Tipunin for interesting discussions.  This work
is supported in part by Deutsche Forschungsgemeinschaft under contract
436 RUS 113-29, and by the RFFI grant 96-01-00725.


\small
\begin{thebibliography}{33}
\parindent=0pt
\parskip=-2pt

\bibitem{[Ade]} M.~Ademollo, L.~Brink, A.~D'Adda, R.~D'Auria, E.~Napolitano,
S.~Sciuto, E.~Del~Guidice, P.~Di~Vecchia, S.~Ferrara, F.~Gliozzi, R.~Musto,
and R.~Pettorino, \PLB62 (1976) 105;\\ M.~Ademollo, L.~Brink, A.~D'Adda,
R.~D'Auria, E.~Napolitano, S.~Sciuto, E.~Del~Guidice, P.~Di~Vecchia,
S.~Ferrara, F.~Gliozzi, R.~Musto, R.~Pettorino, and J.H.~Schwarz, \NPB111
(1976) 77.

\bibitem{[AGSY]} O.~Aharony, O.~Ganor, J.~Sonnenschein, S.~Yankielowicz,
and N.~Sochen,
\NPB399 (1993) 527;\\
O.~Aharony, J.~Sonnenschein, and S.~Yankielowicz,
\PLB289 (1992) 309.


\bibitem{[Andreev]} O.~Andreev, 
\PLB363 (1995) 166.

\bibitem{[AY]} H.~Awata and Y.~Yamada, \MPLA7 (1992) 1185.

\bibitem{[BBS]} O.~Babelon, D.~Bernard, and F.A.~Smirnov,
{\it Null Vectors in Integrable Field Theory\/},
SACLAY-SPHT-96-063, hep-th/9606068.

\bibitem{[B]} Z.~Bajnok, \PLB320 (1994) 36; \PLB329 (1994) 225.

\bibitem{[BOP]}F.~Bastianelli, N.~Ohta and J.L.~Petersen, \PLB327 (1994) 35.

\bibitem{[BdFIZ]} M.~Bauer, P.~di~Francesco, C.~Itzykson, and J.-B.~Zuber,
\NPB362 (1991) 515.

\bibitem{[BS]} M.~Bauer and N.~Sochen, \CMP\ 152 (1993) 127.

\bibitem{[BPZ]} A.A.~Belavin, A.M.~Polyakov, and A.B.~Zamolodchikov, \NPB241
(1984) 333.

\bibitem{[BSAVir]} L.~Benoit and Y.~Saint-Aubin, \PLB215 (1987) 517; \
Lett. Math. Phys. 23 (1991) 117.

\bibitem{[BOh]}N.~Berkovits and N.~Ohta,
\PLB334 (1994) 72.

\bibitem{[BV]}N.~Berkovits and C.~Vafa, \MPLA9 (1994) 653.

\bibitem{[BV-top]}N.~Berkovits and C.~Vafa, \NPB433 (1995) 123.

\bibitem{[BLNW]}M.~Bershadsky, W.~Lerche, D.~Nemeschansky, and N.P.~Warner,
Nucl. Phys. B401 (1993) 304.

\bibitem{[BO]}M.~Bershadsky and H.~Ooguri, \PLB229 (1989) 374.

\bibitem{[BLLS]} A.~Boresch, K.~Landsteiner, W.~Lerche and A.~Sevrin, \NPB436
(1995) 609.

\bibitem{[BFK]} W.~Boucher, D.~Friedan, and A.~Kent, \PLB172 (1986) 316.

\bibitem{[BMP]} P.~Bouwknegt, J.~McCarthy, and K.~Pilch,
J.\ Geom.\ Phys.\ 11 (1993) 225.

\bibitem{[BKT]}P.~Bowcock, R.L.~Koktava, and A.~Taormina, {\it Free field
Representations for the Affine Superalgebra $\widehat{sl(2|1)}$
and noncritical $\N2$ strings\/}, hep-th/9606015.

\bibitem{[BT]}P.~Bowcock and A.~Taormina, {\it Representation theory
of the affine Lie superalgebra $sl(2|1)$ at fractional level\/},
hep-th/9605220.

\bibitem{[BWW3]} P.~Bowcock and G.M.T.~Watts, \PLB297 (1992) 282.

\bibitem{[Doerr2]}M.~D\"orrzapf, {\it Analytic Expressions for the Singular
Vectors of the $N=2$ Superconformal Algebra\/},\\ hep-th/9601056.

\bibitem{[DF]} Vl.S.~Dotsenko and V.A.~Fateev, \NPB 240 (1984) 312.

\bibitem{[Ey]}T.~Eguchi and S.-K.~Yang, \MPLA4 (1990) 1653.

\bibitem{[FY]}J.-B.~Fan and M.~Yu, {\it Modules over Affine Lie
Superalgebras\/}, hep-th/9304122;
{\it $G/G$ Gauged Supergroup Valued WZNW Field Theory\/}, hep-th/9304123.

\bibitem{[FF]}B.L.~Feigin and D.B.~Fuchs, Funct. Anal. Appl. 16 (1982) 114.

\bibitem{[FM]}B.~Feigin and F.~Malikov, {\it Integral intertwining operators
and complex powers of differential ($q$-difference) operators\/},
Kyoto preprint, RIMS-894.

\bibitem{[FS]}
A.V.~Stoianovsky and B.L.~Feigin, 
Funk. An. i ego prilozh., 28(1) (1994) 68;
28(4) (1994) 42.

\bibitem{[FST]}B.L.~Feigin, A.M.~Semikhatov, and I.Yu.~Tipunin,
{\it Equivalence between Categories of Verma Modules over
Affine $\SL2$ and $N=2$ Superconformal Algebras\/}, to appear.

\bibitem{[FoF]}J.M.~Figueroa-O'Farrill, \PLB 321 (1994) 344; \NPB432 (1994)
404.

\bibitem{[FT]}  E.S.~Fradkin and A.A.~Tseytlin, \PLB106 (1981) 63; \PLB162
(1985) 295.

\bibitem{[FMS]}D.H.~Friedan, E.J.~Martinec, and S.H.~Shenker, \NPB271 (1986)
93.

\bibitem{[Gd]}M.R.~Gaberdiel, {\it Fusion of twisted representations},
hep-th/9607036.

\bibitem{[GP]} A.Ch.~Ganchev and V.B.~Petkova, \PLB293 (1992) 56; \
\PLB318 (1993) 77.

\bibitem{[GS2]} B.~Gato-Rivera and A.M.~Semikhatov, \PLB293 (1992) 72,
Theor. Math. Phys. 95 (1993) 536.

\bibitem{[GR]} A.~Giveon and M.~Ro\v cek, \NPB400 (1993) 145.

\bibitem{[HY]}H.-L.~Hu and M.~Yu,
\NPB391 (1993) 389.

\bibitem{[IK]} K.~Ito and H.~Kanno,
\MPLA9 (1994) 1377.

\bibitem{[KK]}V.G.~Ka\v{c} and D.A.~Kazhdan, Adv. Math. 34 (1979) 97.

\bibitem{[KS]}H.~Kanno and M.H.~Sarmadi,
Int.\ J.\ Mod.\ Phys.\ A9 (1994) 39.

\bibitem{[K1]} A.~Kent, \PLB273 (1991) 56.

\bibitem{[Ketov]} S.V.~Ketov, {\it The $Osp(32|1)$ versus $Osp(8|2)$
supersymmetric M-brane action from self-dual $(2,2)$ strings\/},
hep-th/9609004.

\bibitem{[KL]} S.V.~Ketov and O.~Lechtenfeld, {\it The String Measure
and Spectral Flow of Critical $N=2$ strings\/}, \PLB353 (1995)
463--470.

\bibitem{[KM]}D.~Kutasov and E.~Martinec, {\it New Principles for
String/Membrane Unification\/}, hep-th/9602049;\\ D.~Kutasov and
E.~Martinec, and M.~O'Loughlin, {\it Vacua of M-theory and N=2
strings\/}, hep-th/9603116.

\bibitem{[LLS]} K.~Landsteiner, W.~Lerche and A.~Sevrin,  \PLB352 (1995) 286.

\bibitem{[Lechtenfeld]}O.~Lechtenfeld, {\it Integration Measure and
Spectral Flow in the Critical $\N2$ String\/}, hep-th/9512189.

\bibitem{[LVW]}W.~Lerche, C.~Vafa, and N.P.~Warner, \NPB324 (1989) 427.

\bibitem{[LZ]}B.H.~Lian and G.J.~Zuckerman, \PLB254 (1991) 417.

\bibitem{[MFF]} F.G.~Malikov, B.L.~Feigin, and D.B.~Fuchs, Funk.\ An.\
Prilozh.\ 20 N2 (1986) 25.

\bibitem{[Marcus]}N.~Marcus, {\it A tour through $N=2$ strings\/}, talk at the
Rome String Theory Workshop, 1992, hep-th/9211059.

\bibitem{[Mart]} E.~Martinec, {\it Geometrical Structures
of M-Theory\/}, EFI-96-29.

\bibitem{[MW]}P.~Mathieu and G.~Watts,
\NPB475 (1996) 361-396. 

\bibitem{[MM]}S.~Mathur and S.~Mukhi, \PRD36 (1987) 465; \NPB302 (1988) 130.

\bibitem{[IMM]}S.~Mukhi, {\it Extra States in $C<1$ String Theory\/},
(Talk given at Cargese Summer School, July 1991), hep-th/9111013 [abs,
src, ps, other];\\
C. Imbimbo, S. Mahapatra and S. Mukhi, \NPB375 (1992) 399

\bibitem{[OP]} N.~Ohta and J.L.~Petersen, \PLB325 (1994) 67.

\bibitem{[OV23]}H.~Ooguri and C.~Vafa \NPB361 (1991) 469; \NPB367 (1991) 83.

\bibitem{[OV-top]}H.~Ooguri and C.~Vafa \NPB451 (1995) 212.

\bibitem{[Petersen]}J.L.~Petersen, J.~Rasmussen, and M.~Yu, \NPB457
(1995) 309.

\bibitem{[RSS]}E.~Ragoucy, A.~Sevrin and P.~Sorba, 
Commun.\ Math.\ Phys. 181 (1996) 91--129.

\bibitem{[SS]}A.~Schwimmer and N.~Seiberg, \PLB184 (1987) 191.

\bibitem{[S-sing]} A.M.~Semikhatov,
\MPLA9 (1994) 1867.

\bibitem{[S-inv]}A.M.~Semikhatov, {\it Inverting the Hamiltonian
Reduction in String Theory\/}, Talk at the 28th Symposium on the
Theory of Elementary Particles, Wendisch-Rietz, September 1994,
hep-th/9410109.

\bibitem{[S-sl21]} A.M.~Semikhatov, {\it The Non-Critical $N=2$ String is an
$\SSL21$ Theory\/}, {\tt hep-th/9604105}, \NPB, to appear.

\bibitem{[ST1]}A.M.~Semikhatov and I.Yu.~Tipunin, \IJMPA 11 (1996) 2721.

\bibitem{[ST2]}A.M.~Semikhatov and I.Yu.~Tipunin, 
\IJMPA11 (1996) 4597.

\bibitem{[ST3]}A.M.~Semikhatov and I.Yu.~Tipunin, {\it All Singular
Vectors of the $N\!=\!2$ Superconformal Algebra via the Algebraic
Continuation Approach\/}, hep-th/9604176.


\bibitem{[W1]}G.M.T.~Watts, \NPB407 (1993) 213.

\bibitem{[W-top]}E.~Witten, Commun. Math. Phys. 118 (1988) 411; \NPB 340
(1990) 281.

\end{thebibliography}
\end{document}